\documentclass[11pt]{article}
 \usepackage{mapletab,mapleenv}

\usepackage{epsfig }
\usepackage{latexsym}
\usepackage{amsfonts,amsmath,amssymb,textcomp}
\newtheorem{thm}{Theorem}[section]
\newtheorem{lem}[thm]{Lemma}
\newtheorem{cor}[thm]{Corollary}
\newtheorem{prop}[thm]{Proposition}
\newtheorem{rem}[thm]{Remark}
\newtheorem{deff}[thm]{Definition}
\newtheorem{conj}[thm]{Conjecture}
\newtheorem{key}[thm]{Keywords}
\newtheorem{prob}[thm]{Problem}
\newenvironment{probb}{ \begin{prob} \rm}{ \end{prob} }
\newcommand{\bprobb}{\begin{probb}}
\newcommand{\eprobb}{\end{probb}}
\newcommand{\bth}{\begin{thm}}
\newcommand{\ethGL}{\end{thm}}
\newcommand{\bconj}{\begin{conj}}
\newcommand{\econj}{\end{conj}}
\newcommand{\bkey}{\begin{key}}
\newcommand{\ekey}{\end{key}}
\newcommand{\bl}{\begin{lem}}
\newcommand{\el}{\end{lem}}
\newcommand{\bdeff}{\begin{deff}}
\newcommand{\edeff}{\end{deff}}
\newcommand{\bcor}{\begin{cor}}
\newcommand{\ecor}{\end{cor}}
\newcommand{\bprop}{\begin{prop}}
\newcommand{\eprop}{\end{prop}}
\newcommand{\brem}{\begin{rem}}
\newcommand{\erem}{\end{rem}}
\newcommand{\beq}{\begin{equation}}
\newcommand{\eeq}{\end{equation}}
\newcommand{\beqn}{\begin{eqnarray}}
\newcommand{\eeqn}{\end{eqnarray}}
\newcommand{\beqns}{\begin{eqnarray*}}
\newcommand{\eeqns}{\end{eqnarray*}}
\newcommand{\ba}{\begin{array}}
\newcommand{\ea}{\end{array}}
\newcommand{\bit}{\begin{itemize}}
\newcommand{\eit}{\end{itemize}}
\newcommand{\ben}{\begin{enumerate}}
\newcommand{\een}{\end{enumerate}}

\newcommand{\BO}{\mathcal{O}}

\newcommand{\non}{\nonumber}
\newcommand{\babs}{\begin{abstract}}
\newcommand{\eabs}{\end{abstract}}
\newcommand{\bal}{\begin{align}}
\newcommand{\bals}{\begin{align*}}
\newcommand{\bs}{\begin{skip}}
\newcommand{\eal}{\end{align}}
\newcommand{\eals}{\end{align*}}
\newcommand{\es}{\end{skip}}

\newcommand{\ra}{\rightarrow}

\newcommand{\B}{\quad}

\newcommand{\lp}{\left (}
\newcommand{\rp}{\right )}

\newcommand{\js}{j^*}
\newcommand{\ks}{k^*}
 
\newcommand{\Cs}{C^*} 
\newcommand{\Ds}{D^*}

		\newcommand{\qt}{\tilde{q}}

\newcommand{\de}{\delta}

\newcommand{\be}{\beta}

\newcommand{\Fi}{\varphi}
\newcommand{\FI}{\phi}
\newcommand{\eps}{\varepsilon}

\newcommand{\gam}{\gamma}

\def\1{{\ifmmode 1\mskip-1.5\thinmuskip\mathrm{l}%
        \else\textrm{1\hskip -.23em l}\fi}}

\def\P{{\mathbb {P}}}

\newcommand{\bin}[2]
{
{#1\choose #2}
}

\title{\bf A refined and asymptotic analysis of optimal stopping problems of Bruss and  Weber    }

\author{Guy Louchard\thanks{Universit\'e Libre de Bruxelles,
D\'epartement d'Informatique, CP 212, Boulevard du Triomphe, B-1050
Bruxelles, Belgium, email: louchard@ulb.ac.be}
}

\date{\today}
\begin{document}

\maketitle

\babs

The  classical secretary problem 
has been generalized over the years into several directions. In this paper we confine our interest to those generalizations which have to do with the more general problem of stopping on a last observation of a specific kind. We follow Dendievel \cite{DE15}, 
\cite{DE16}, (where a  bibliography can be found) who studies several types of such problems, mainly initiated by Bruss \cite{Br841} and \cite{BRU00},    Weber \cite{WE131} and \cite{WE13}. Whether in discrete time or continuous time, whether all parameters are known or must be sequentially estimated, we shall call such problems simply \textit {Bruss-Weber problems.}
Our contribution in the present paper is a refined analysis of several problems in this class and a study of the asymptotic behaviour of solutions. 

The problems we consider center around the following model. Let $X_1,X_2,\ldots,X_n$ be a sequence of  independent random variables which can take three values: $\{+1,-1,0\}.$ 
Let $p:=\P(X_i=1), p':=\P(X_i=-1), \qt:=\P(X_i=0), p\geq p'$, where $p+p'+\qt=1$. The goal is to maximize the probability of stopping on a value $+1$ or $-1$ appearing for the last time in the sequence. Following a suggestion by  Bruss,  we have also analyzed an  x-strategy with incomplete information:  the cases $p$ known, $n$ unknown, then $n$ known, $p$ unknown and finally $n,p$  unknown are considered. We also present simulations of the corresponding complete selection algorithm.
\eabs


\section{Introduction}\label{S1}

The  classical secretary problem 
has been generalized over the years into several directions. In this paper we confine our interest to those generalizations which have to do with the more general problem of stopping on a last observation of a specific kind. We follow Dendievel \cite{DE15}, \cite{DE16}, (where a  bibliography can be found) who studies several types of such problems, mainly initiated by Bruss \cite{Br841}, 
\cite{BRU00}  and   Weber \cite{WE131}, \cite{WE13}. Whether in discrete time or continuous time, whether all parameters are known or must be sequentially estimated, we shall call such problems simply \textit {Bruss-Weber problems.}

Bruss \cite{BRU00} studied the case of stopping on a last  $1$ in a sequence of $n$ independent random variables
 $X_1,X_2,\ldots,X_n$, taking values $\{1,0\}$. This led to the versatile odds-algorithm and also to a similar method in  continuous-time,
allowing for interesting applications in different domains, as e.g. in investment problems studied in Bruss and Ferguson \cite{BF02}.
See also  Szajowski and  {\L}ebek \cite{SL07}. Moreover, Bruss and Louchard \cite{BL08P}
  studied the case
 where the odds are unknown and have to be sequentially estimated, showing a convincing stability for applications.

 Weber   (R.R. Weber, University of Cambridge), considered the model  of iid random variables taking values in  $\{+1,-1,0\}$. The goal is to maximize the probability of stopping on a value $+1$ or $-1$ appearing for the last time in the sequence. The background was as follows.

When teaching  the odds-algorithm in his course (see section 6 of his course on optimization and control \cite{WE131}), Weber   proposed the following problem to his students:

\textit {A financial advisor can impress his clients if immediately following a week in which the FTSE index moves by more than $5\%$ in  some direction he correctly predicts that this is the last week during the calendar year that it moves more than $5\%$ in that direction}

\textit{Suppose that in each week the change in the index is independently up by at least $5\%$, down by at least $5\%$ or neither of these, with probabilities $p$, $p$ and $1-2p$ respectively ($p\leq 1/2$). He makes at most one prediction this year. With what strategy does he maximize the probability of impressing his clients?}

The solution of this interesting problem is easy but can only be partially retrieved from the odds-algorithm.

Weber \cite{WE13} then discussed with Bruss several more difficult versions  of this problem, some of them studied in  Dendievel's PhD thesis \cite {DE16}. 

Let us also mention shortly related work:  Hsiau and Yang \cite{HS02} have studied the problem of stopping on a last $1$ in a sequence of Bernoulli trials in a Markovian framework, where the value taken by the $k$th variable is influenced by the value of the the $(k-1)$th variable.  Ano and Ando  \cite{AA00}, generalizing the model of Bruss \cite{BR87}, consider options arising according to a 
{P}oisson process with unknown intensity but only available with a fixed probability $p$. 
 Tamaki \cite{TA10} generalized the odds-algorithm by introducing multiplicative odds in order to solve the problem of optimal stopping on any of a fixed number of last successes. Surprising coincidences of lower bounds for odds-problems with multiple stopping have been discovered by Matsui and Ano \cite{MA16}, generalizing Bruss \cite{BRU03}. A more specific interesting problem of multiple stopping in Bernoulli trials with a random number of observations was studied by Kurushima and Ano \cite{KA16}.   
  \vspace{1 em}

 Let $p:=\P(X_i=1),p':=\P(X_i=-1),\qt:=\P(X_i=0), p\geq p'$, where $p+p'+\qt=1$.
 
A first problem studied in \cite{DE15} is to maximize for a fixed number $n$ of variables the success probability $w_{j,k},j\geq k$ with the following strategy: we observe $X_1,X_2,\ldots$.
Wait until $i=k$. From $k$ on, if $X_i=-1$ we select $X_i$ and stop. If not we proceed to the next random variable and start the algorithm again. If no $-1$ value was found before $j$, then, from $j$ on, if $X_i=+1$ or $X_i=-1$ we select this variable and stop. If none was found (all $X_i=0$ from $j$ to $n$) then we fail. The goal is to find $j^*,k^*$ such that $w_{j^*,k^*}$ is maximum. 
In \cite{DE15}{}, explicit expressions for $w_{j,k},w_{j,j}$ are given and $j^*,k^*$ are numerically computed for given $n$.  Dendievel also proves that the problem is monotone in the sense of Assaf and Samuel-Cahn \cite{AS00}: if at a certain time it is optimal to stop on a $1$ 
(respectively on a $-1$), then it is optimal to  stop on a $1$(respectively on a $-1$) at any later time index. Also, it is proved in \cite{DE15}, that if $p\geq p'$ then $j^*\geq k\*$.

Our contribution is the following: in Section \ref{S2}, we provide explicit optimal solutions in a continuous model and in the present  discrete case for $p>p'$ and $p=p'$.

Another problem, initiated by a model of Bruss in continuous time, and leading to the 1/e-law of best choice (Bruss \cite {Br841})
is a problem in continuous time, now with a fixed total number of variables $n$ with possible values in ${0,-1,1}$. More precisely, let $U_i,i=1,2,\ldots,n$  be independent random variables uniformly distributed on the  interval $[0,1]$. Let $T_i=U_{\{i\}}$: $T_i$ is the $i$th order statistic of the $U_i$'s. $T_i$ is the arrival time of $X_i$. The strategy is to wait until some time $x_n^*$ and from $x_n^*$ on, we select the 
first $X_i=+1$ or   $X_i=-1$, using the previous algorithm with $p=p'$. Following Bruss \cite{BRU00}, we call this strategy an x-strategy. In \cite{DE15}, for this problem, the author gives  the optimal $x_n^*$ and  the corresponding success probability $P_n^*$.

In Section \ref{S3} we provide some asymptotic expansions for this x-strategy's parameters,   for $p=p'$ . We also consider the  success probability for small $p$ and for the case $p>p'$.

In Section \ref{S4}, following a suggestion by  Bruss,  we have analyzed an  x-strategy with incomplete information:  the cases $p$ known, $n$ unknown, then $n$ known, $p$ unknown and finally $n,p$  unknown are considered. We also present simulations of the complete selection algorithm. 

%
\section{The optimal solution}\label{S2}
In this Section, we analyze explicitly the optimal solutions in the continuous and discrete case for $p>p'$ and $p=p'$.
The following notations will be used in the sequel: $q:=1-p,q'=1-p',\qt=1-p-p'$.
\subsection{The optimal solution, continuous case, $p>p'$}

Let us first consider $p>p',j\ge k$.
The success probabilities satisfy the following forward recurrence equations 
(these are easily obtained from the stopping times characterizations):
\bal
w_{j,j}&=pq^{n-j}+p'q'^{n - j}+\qt w_{j+1,j+1},\B w_{n,n}=p+p',                         \label{E1}\\
w_{j,k} &=p'q'^{n - k}+q'w_{j,k+1}.                                                         \label{E2}
\end{align}
The solutions, already given in Dendievel \cite{DE15}, are

\beq
{w_{jj}} = (p^{2}\,q^{n - j + 1} - p^{2}\,\tilde{q}^{n-j+1} + p'^{2}\,q'^{
n-j+1} 
\mbox{} - p'^{2}\,\tilde{q}^{n-j+1})/(p'\,p) ,                                     \label{E3}
 \eeq

\beq
{w_{j,k}} = (j-k)\,p'\,q'^{n-k} 
\mbox{} + q'^{j-k}\,\lp {\displaystyle \frac {p\,(q^{n-j+1}  
- {\tilde{q}}^{n-j+1})}{p'}}  + 
{\displaystyle \frac {p'\,(q'^{n-j+1} - 
{\tilde{q}}^{n-j+1})}{p}} \rp      .                                        \label{E4}
\eeq
If $j \leq k$, we use
{%
\maplemultiline{
\mathit{w_{k,j}} := (k-j)\,p\,q^{n-j} \\
\mbox{} + q^{k-j}\,\lp{\displaystyle \frac {\mathit{p'}\,(
\mathit{q'}^{n-k+1} - \mathit{\qt}^{n-k+1})}{p}}  + 
{\displaystyle \frac {p\,(q^{n-k+1} - \mathit{\qt}^{n - k
 + 1)}}{\mathit{p'}}} \rp .}
}

 \vspace{1 em}
 \textbf{ Simplification using generating functions}
\vspace{1 em}

We shall show that  these expressions can be nicely derived  by using backward generating functions. Let $F(z):=\sum_{j=-\infty}^{n-1}z^{n-j}w_{j,j}$.
From (\ref{E1}), we have

\[
 {F(z)} - p - p' - {\displaystyle 
\frac {p'\,q'\,z}{1 - z + p'\,z}
}  - {\displaystyle \frac {p\,q\,z}{1 - z + z\,p}} 
 - \tilde{q}\,z\,{F(z)}=0,
\]
the solution of which is

\bals
F(z)&= {\displaystyle \frac { - p'\,z + p + 
p' + 2\,p\,p'\,z - z\,p}{(1 - z + z\,p)\,(1 - z
 + p'\,z)\,(1 - z + z\,p + p'\,z)}} \\
&{\displaystyle =-\frac {(p^{2} + p'^{2})\,\tilde{q}}{p'\,p\,(1 - z + z\,p + p'\,z)}} 
 + {\displaystyle \frac {p'\,q'}{p\,(1
 - z + p'\,z)}}  + {\displaystyle \frac {p\,q}{
p'\,(1 - z + z\,p)}} .
\end{align*}
This immediately leads to (\ref{E3}). Similarly, let $F_j(z):=\sum_{k=-\infty}^{j-1}z^{j-k}w_{j,k}$. From (\ref{E2}) this satisfies

\bals
& F_j(z) - \lp p^{2}\,q^{n-j+1} - p
^{2}\,\tilde{q}^{n-j+1} + p'^{2}\,q'^{n-j+1}\right. \\
&\left. \mbox{} - p'^{2}\,\tilde{q}^{n-j+1}\rp/(
p'\,p)\mbox{} - {\displaystyle \frac {p'\,z}{q^{' - n + j - 1}\,(1 - z + p'\,z)}}  
\mbox{} - q'\,z\,F_j(z) =0,
\end{align*}
the solution of which, expanded into partial fractions,  leads to 

{%
\maplemultiline{
 F_j(z)= \lp - p'^{3}\,q'^{n-j
} + p'^{3}\,\tilde{q}^{n-j} - p\,
p'^{2}\,q'^{n-j}
\mbox{} + p'^{2}\,q'^{n-j} + p\,
p'^{2}\,\tilde{q}^{n-j} - p'^{2
}\,\tilde{q}^{n-j} \right.\\
\left. \mbox{} + p'\,p^{2}\,\tilde{q}^{n-j} - p
^{2}\,\tilde{q}^{n-j} + p^{2}\,q^{n-j}
\mbox{} - p^{3}\,q^{n-j} + p^{3}\,\tilde{q}
^{n-j}\rp/((1 - z + p'\,z)\,p'\,p)\mbox{} + 
{\displaystyle \frac {p'\,q'^{n-j}}{(
1 - z + p'\,z)^{2}}} . }
}
This simplifies as
\[ F_j(z)=\lp p^2qq^{n-j}+p'^{2}\qt q'^{n-j} -(p^2+p'^{2})\qt^{n-j+1}\rp/((1-z+p'z)pp') +{\displaystyle \frac {p'\,q'^{n-j}}{(
1 - z + p'\,z)^{2}}}. \]
Now from  (\ref{E2}) the presumed generating function is given by 

\bals
Fth_j(z) &=  \frac {p'\,z}{q'^{( - n + j - 1)}\,( - 1 + q'\,z)^{2}}\\
& - q' \left(  
 \frac {p}{p'}\left({\displaystyle \frac {1}{q^{( - n + j - 1)}}} 
 - {\displaystyle \frac {1}{\tilde{q}^{( - n + j - 1)}}}\rp  
 \mbox{} +\frac {p'}{p}\left({\displaystyle \frac {1}{q'^{( - n + j - 1)}}} 
 - {\displaystyle \frac {1}{\tilde{q}^{( - n + j - 1)}}}\rp  
 \right)\frac{z}{( - 1 + q'\,z)}.
\end{align*}
Identification with $F_j(z)$ is immediate.

\vspace{1 em}
 \textbf{ Computation of the optimal values $\js,\ks$}
\vspace{1 em}

Let us now turn to the main object of this Section which is  the computation of the optimal values $\js,\ks$. It is proved in \cite{DE15} that, if $p>p'$ then $\js\ge \js$. Actually, setting $j=n-C,k=n-D$ in (\ref{E3}),(\ref{E4}), we see that $w_{j,k},w_{k,j}$  \textit{ do not depend on $n$ }. We have, with 
$C\leq D$, and using $C,D$ as continuous variables,

\[
{w_{C,D}} := ( - {C} + {D})\,p'\,
q'^{{D}}
\mbox{} + q'^{ - {C} + {D}}\,\lp
{\displaystyle \frac {p\,(q^{{C} + 1} - {\tilde{q}}^{{C} + 1})}p'}  + {\displaystyle \frac {
p'\,(q'^{C+1} - {\tilde{q}}^{{C} + 1)}}{p}} \rp ,\]
and if $D\leq C$,
{%
\maplemultiline{
{w_{D,C}} := ( - {D} + {C})\,p\,q^{{
C}}
\mbox{} + q^{ - {D} + {C}}\,\lp{\displaystyle 
\frac {p'\,(q'^{{D} + 1} - {\tilde{q}
}^{{D} + 1})}{p}}  + {\displaystyle \frac {p\,(q^{{D} + 1} - {\tilde{q}}^{{D} + 1})}{p'
}} \rp .}
}

The optimal value $C^*$ is the (unique) solution of 
\bal
&\FI_1(\Cs)=0,                                                                    \label{E50}\\   
&\FI_1(C):=\frac{\partial w_{C,D}}{\partial C}q'^{C-D}pp'=
 - {\tilde{q}}\,(p^{2} + p'^{2})\,( - \mathrm{ln}(
q') + \mathrm{ln}({\tilde{q}}))\qt^C+p^{2}\,q\,( - \mathrm{ln}(q') + \mathrm{ln}(q))q^C-p'^{2}pq'^{C}.    \label{E5}
\end{align}

\begin{figure}[htbp]
	\centering
		\includegraphics[width=0.8\textwidth,angle=0]{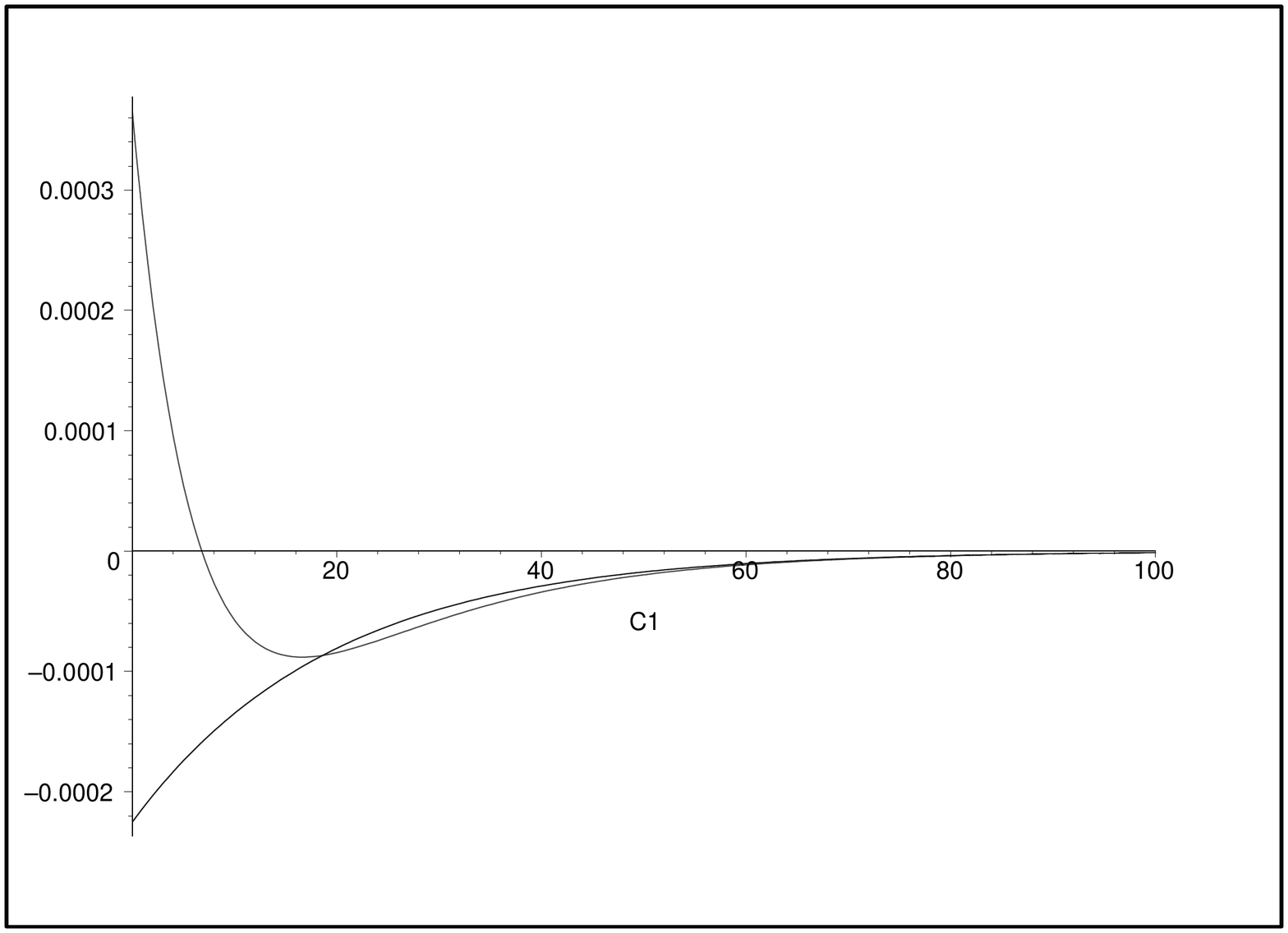}
	\caption{$\FI_1(C)$, $p=0.09,p'=0.05$, together with $\FI_{as}(C)$ (lower curve)  }
	\label{F1}
\end{figure}

First of all, we have $\qt<q<q'$ , $p'<p$ for $0\leq p\leq 1/2$, $p'<1-p$ for $1/2 \leq p\leq 1$. Dividing Eq. (\ref{E5}) by $q'^{C}$, we see that $\FI_1(C)\sim \FI_{as}(C)= -p'^{2}pq'^{C}, C\ra  \infty$ which is negative. 
A plot of $\FI_1(C)$ , for $p=0.09,p'=0.05$ is given in Figure \ref{F1}, together with $\FI_{as}(C)$,  showing numerically a unique maximum, but   we need a formal proof.

\begin{figure}[htbp]
	\centering
		\includegraphics[width=0.8\textwidth,angle=0]{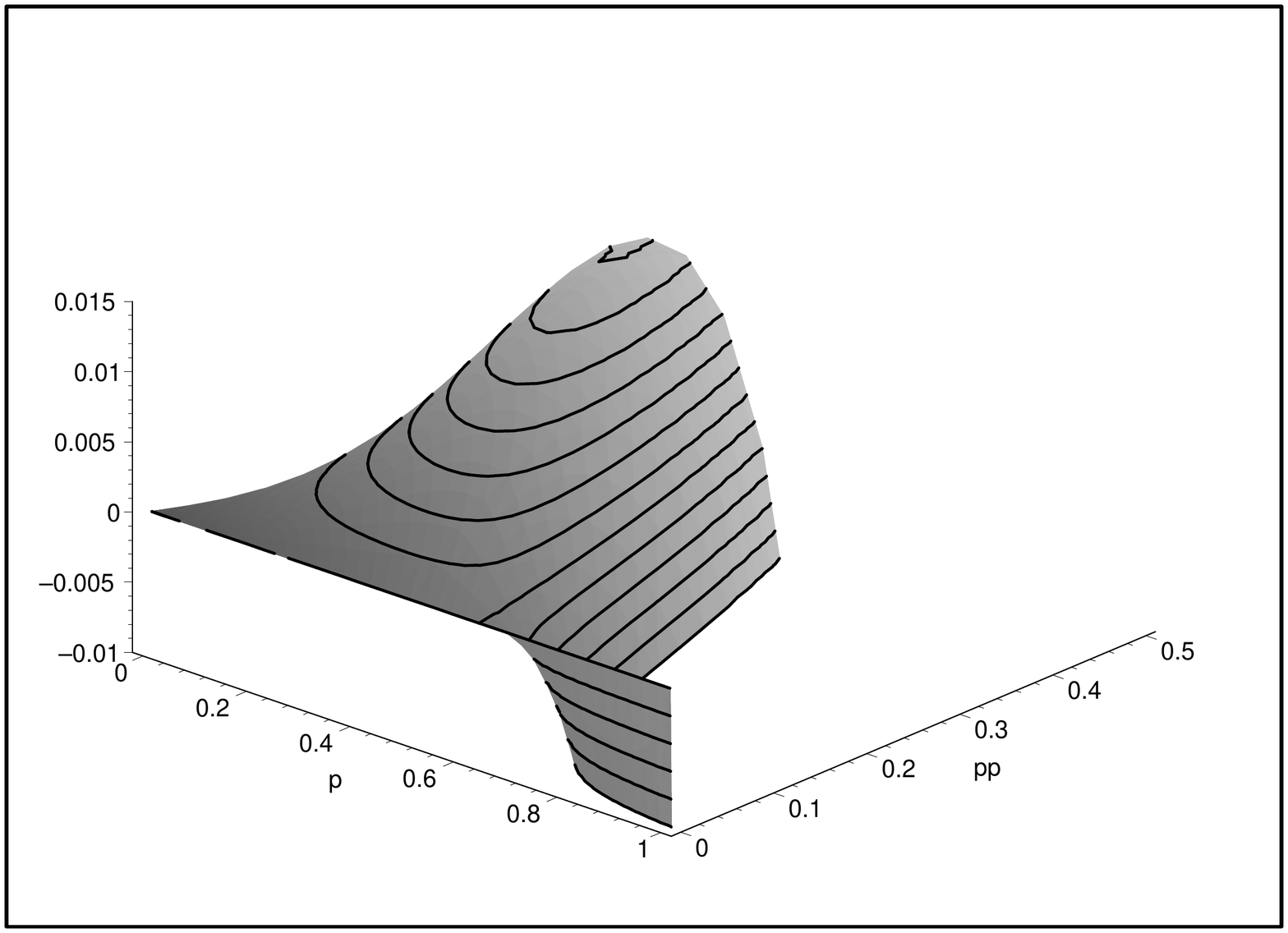}
	\caption{A plot $\FI_1(0)$ defined in  Eq.\ref{E5} as a function of $p,p'$}
	\label{F31}
\end{figure}

\begin{figure}[htbp]
	\centering
		\includegraphics[width=0.8\textwidth,angle=0]{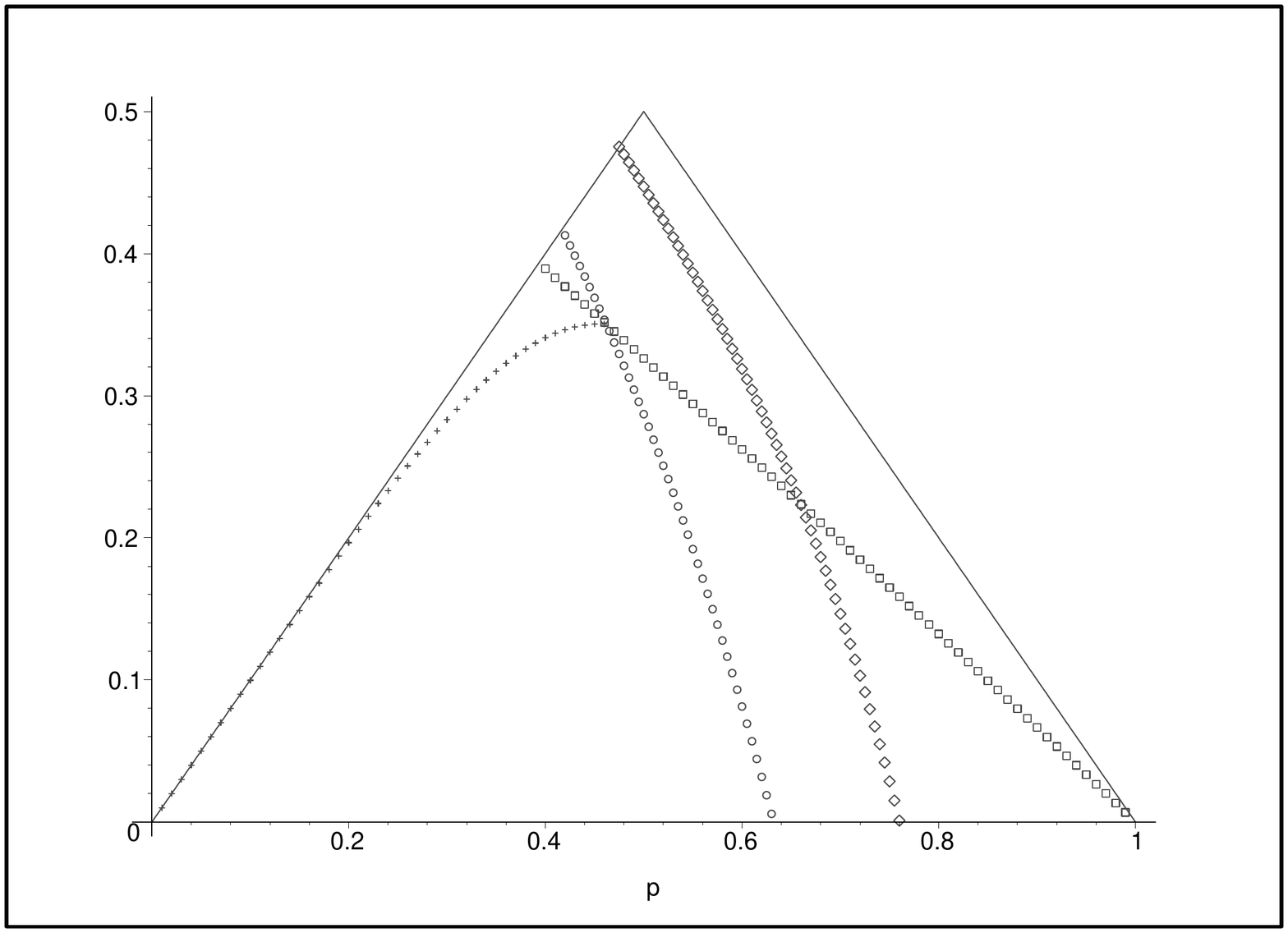}
	\caption{ The graphic  shows the functions  $\gam_1(p)$  (circles), $\gam_2(p)$  (box), $\gam_3(p)$ (cross), $\gam_4(p)$ (diamonds) defined in the text  with the constraints on $(p,p')$ }
	\label{F32}
\end{figure}

We would like to have $\FI_1(0)>0$, this would imply the existence of  $C^*$. A plot of 
$\FI_1(0)$ (satisfying the constraints on $(p,p'))$ is given in  Figure \ref{F31}. We see that there exists a curve $p'= \gam_1(p)$, given in Figure \ref{F32}, such that $\FI_1(0)<0$ if $p'> \gam_1(p)$. In this case, we must choose $C^*=0$. Otherwise, we know that $C^*$
does exist. The extremal points of $\gam_1(p)$  are $(0.4170224307\ldots,0.4170224307\ldots),(0.63212005588\ldots,0)$.

Finally, we must prove the uniqueness  of $C^*$. By dividing  Eq.(\ref{E50}) by $q'^{C}$, we obtain,
 with $\tilde{r}:=\qt/q',r:=q/q',\tilde{r}<r$,
\[A_1 \tilde{r}^C=A_2 r^C +A_3,\]
where $A_1,A_2,A_3$ do not depend on $C$. On both sides, we have strictly convex/concave functions of $C$ which ensure the uniqueness of
 $C^*$.  

Interestingly, $C^*$ does not depend on $D$. 
The optimal value $D^*$ is the  solution, for $C=\Cs$,  of 

\bals
\frac{\partial w_{C,D}}{\partial D}q'^{-D}pp'&= p'^{2}\,p - p'^{2}\,
\mathrm{ln}(q')\,{C}\,p + p'^{2}\,
\mathrm{ln}(q')\,{D}\,p + \mathrm{ln}({q'})\,q'^{ - {C}}\,p^{2}\,q^{C+1}
 \\
&\mbox{} - \mathrm{ln}(q')\,q'^{ - {C}}\,p^{2}\,{\tilde{q}}^{C+1} + p'^{2}\,
q'\,\mathrm{ln}(q') - \mathrm{ln}(q')
\,q'^{-C}\,p'^{2}\,{\tilde{q}}^{{C} + 1}=0 ,
\end{align*}

this gives
\beq
D=\FI_2(C):= = \lp - {\displaystyle \frac {p\,q\,q^C}{p^{'2}}}+{\displaystyle \frac {{\tilde{q}}\,(p^{2} + p^{'2})}{
p'^{2}\,p}}  \qt^{C} \rp
\,q'^{-C} +  {\displaystyle \frac { - p + \mathrm{ln}(q')\,{C
}\,p - q'\,\mathrm{ln}(q')}{\mathrm{ln}(
q')\,p}},                                                                            \label{E6}
  \eeq
and $D^*=\FI_2(C^*)$.
\newpage
 \vspace{1 em}
 \textbf{ The  acceptance regions}
\vspace{1 em}
\ben
 \item Curiously enough, even if we must choose $C^*=0$ (see above), $D^*$ \textit{is not necessarily non-negative!}
If we solve $\FI_2(0)=0$  w.r.t $p'$ for each $p$, we obtain a second curve $p'= \gam_2(p)$ also given in Figure \ref{F32}.
The extremal points of $\gam_2(p)$  are $(0.3934693403\ldots,0.3934693403\ldots),(1,0)$. If $p'>\gam_2(p)$, then we must choose $D^*=0$ which means waiting until $X_n$. Notice that the two curves do cross.

\item Even more interesting, even if $C^*>0$, $D^*$ \textit{is not necessarily $>C^*$}. If we solve $\{\FI_1(C^*)=0,\FI_2(C^*)=C^*\}$  w.r.t. $\{C^*,p'\}$, we obtain a third curve $p'= \gam_3(p)$ also given in Figure \ref{F32}. If $p'> \gam_3(p)$, we must choose the optimal point on the diagonal: see   the remark below at the end of Section \ref{S23}. The intersection of $\gam_1,\gam_2,\gam_3$ is given by $p_\bullet=0.461926509410\ldots,p'_\bullet=0.350346565861\ldots$.
\item Finally, if we stay above the curve  $\gam_2(p)$, we obtain $\Cs<0$. For instance, for $\Cs=-0.3$, if we solve $\FI_1(-0.3)=0$  w.r.t $p'$ for each $p$, we obtain a fourth  curve $p'= \gam_4(p)$ also given in Figure \ref{F32}. The extremal points of $\gam_4(p)$  are $(0.4751561101\ldots,0.4751561101\ldots),(0.7603489635,0)$. $\gam_4(p)$ is of course not practically useful in our analysis
( we must have $\Cs\geq 0$ ), but it has some interesting asymptotic properties that we detail in Appendix \ref{S6}.
\een 
\newpage
 \vspace{1 em}
 \textbf{ A useful table summarizing acceptance regions}
\vspace{1 em}

The following table \ref{T1} shows the different $\{p,p'\}$ regions and their corresponding $\Cs,\Ds$ characteristics.
 \begin{table}[ht]
 \begin {center}
 \begin{tabular} {|c|c|c|}
\hline
   $p,p'$       &   Theoretical $\Cs,\Ds$       &     Practical    $\Cs,\Ds$    \\
\hline
\hline

 $p'>\gam_1(p), p'>\gam_2(p)$        &    $\Cs<0,\FI_2(0)<0$      &    $\Cs=0,\Ds=0$    \\

  \hline 
	
    $p'=\gam_2(p), p>p_\bullet$      &      $\Cs<0,\FI_2(0)=0$     &      $\Cs=0,\Ds=0$   \\

	\hline 
	
     $p'>\gam_1(p),p'<\gam_2(p), p>p_\bullet$      &     $\Cs<0,\FI_2(0)>0$       &   $\Cs=0,\Ds=\FI_2(0)$      \\

	\hline
	\hline
	
     $p'=\gam_1(p), p<p_\bullet$       &    $\Cs=0,\FI_2(0)<0$       &    $\Cs=0,\Ds=0$     \\

  \hline

      $ p=p_\bullet,p'=p'_\bullet$      &      $\Cs=0,\FI_2(0)=0$       &    $\Cs=0,\Ds=0$     \\

	\hline

      $p'=\gam_1(p), p>p_\bullet$    &     $\Cs=0,\FI_2(0)>0$      &    $\Cs=0,\Ds=\FI_2(0)$     \\

	\hline
	\hline
	
       $p'>\gam_2(p),p'<\gam_1(p), p<p_\bullet$    &     $\Cs>0,\FI_2(0)<0$       &    $\Cs,\Ds=\Cs$     \\

  \hline 
	
     $p'=\gam_2(p),p'>\gam_3(p), p<p_\bullet$    &       $\Cs>0,\FI_2(0)=0$      &    $\Cs,\Ds=\Cs$      \\

	\hline 
	
      $p'<\gam_2(p),p'>\gam_3(p), p<p_\bullet$     &    $\Cs>0,\FI_2(0)>0,\FI_2(\Cs)<\Cs$       &    $\Cs,\Ds=\Cs$   
\footnotemark			\\

	\hline
	
      $p'<\gam_1(p),p'<\gam_2(p), p'<\gam_3(p)$     &     $\Cs>0,\FI_2(\Cs)>\Cs$       &      $\Cs,\Ds=\FI_2(\Cs)$     \\

	\hline
	\hline
\end{tabular} 
\end{center} 
\caption{$\{p,p'\}$ regions and their corresponding $\Cs,\Ds$ characteristics} 
\label{T1} 
\end{table} 

\begin{figure}[htbp] 
	\centering	
		\includegraphics[width=0.8\textwidth,angle=0]{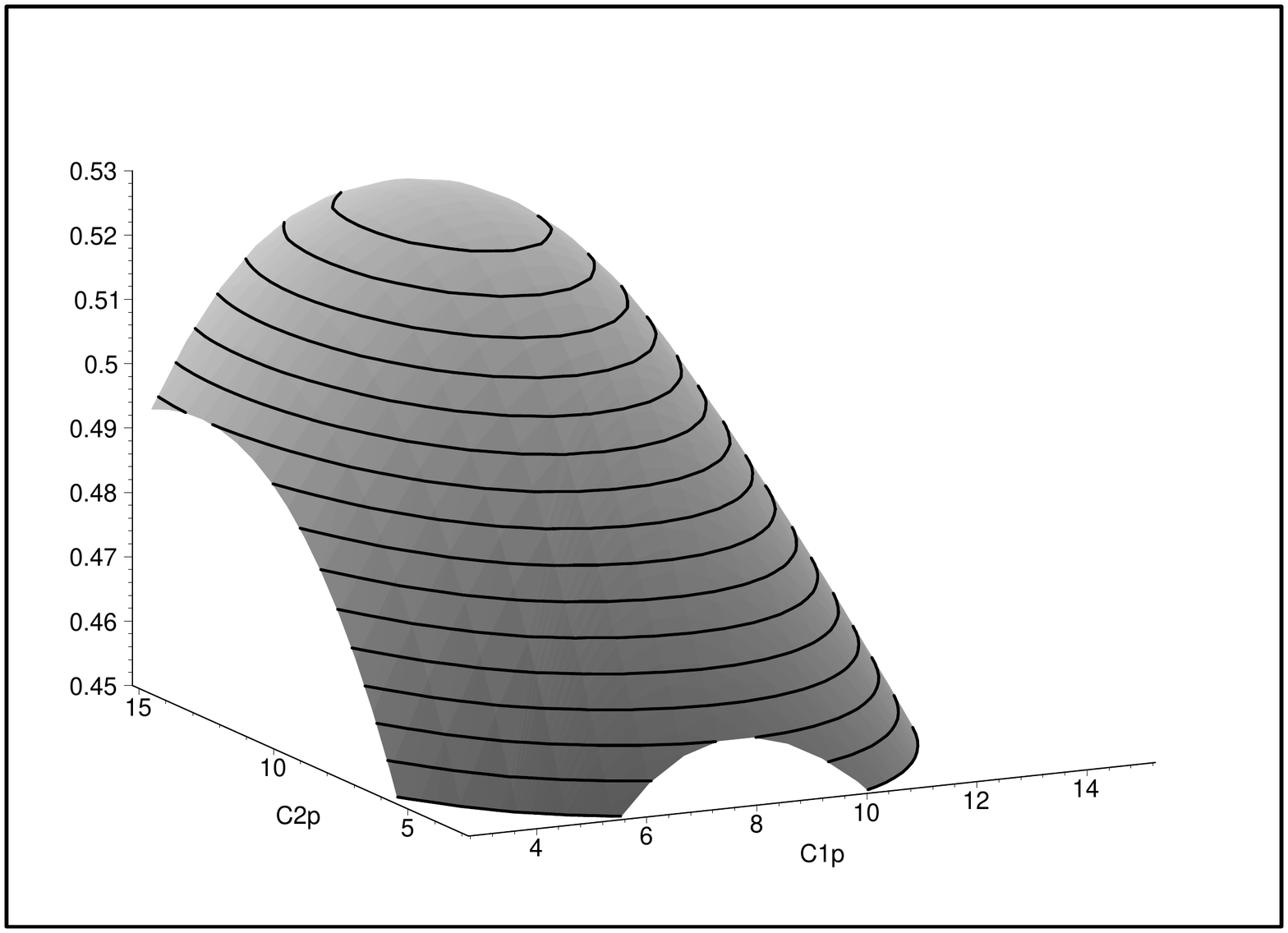} 
			\caption{$w_{C,D},C\leq D$, $w_{D,C},C\geq D$, $p=0.09,p'=0.05$ } 
				\label{F5} 
				\end{figure}

\footnotetext{see   the remark below at the end of Section \ref{S23}}
 As an illustration of the last line of Table \ref{T1}, a plot of $w_{C,D},p=0.09,p'=0.05,C\leq D$ is given in Figure \ref{F5} as well as $w_{D,C},C\geq D$. Also $w_{C^*,D^*} = 0.529979034749\ldots,p=0.09,p'=0.05$. 

\newpage
\subsection{The optimal solution in the  discrete case for  $p>p'$}

\begin{figure}[htbp]
	\centering
		\includegraphics[width=0.8\textwidth,angle=0]{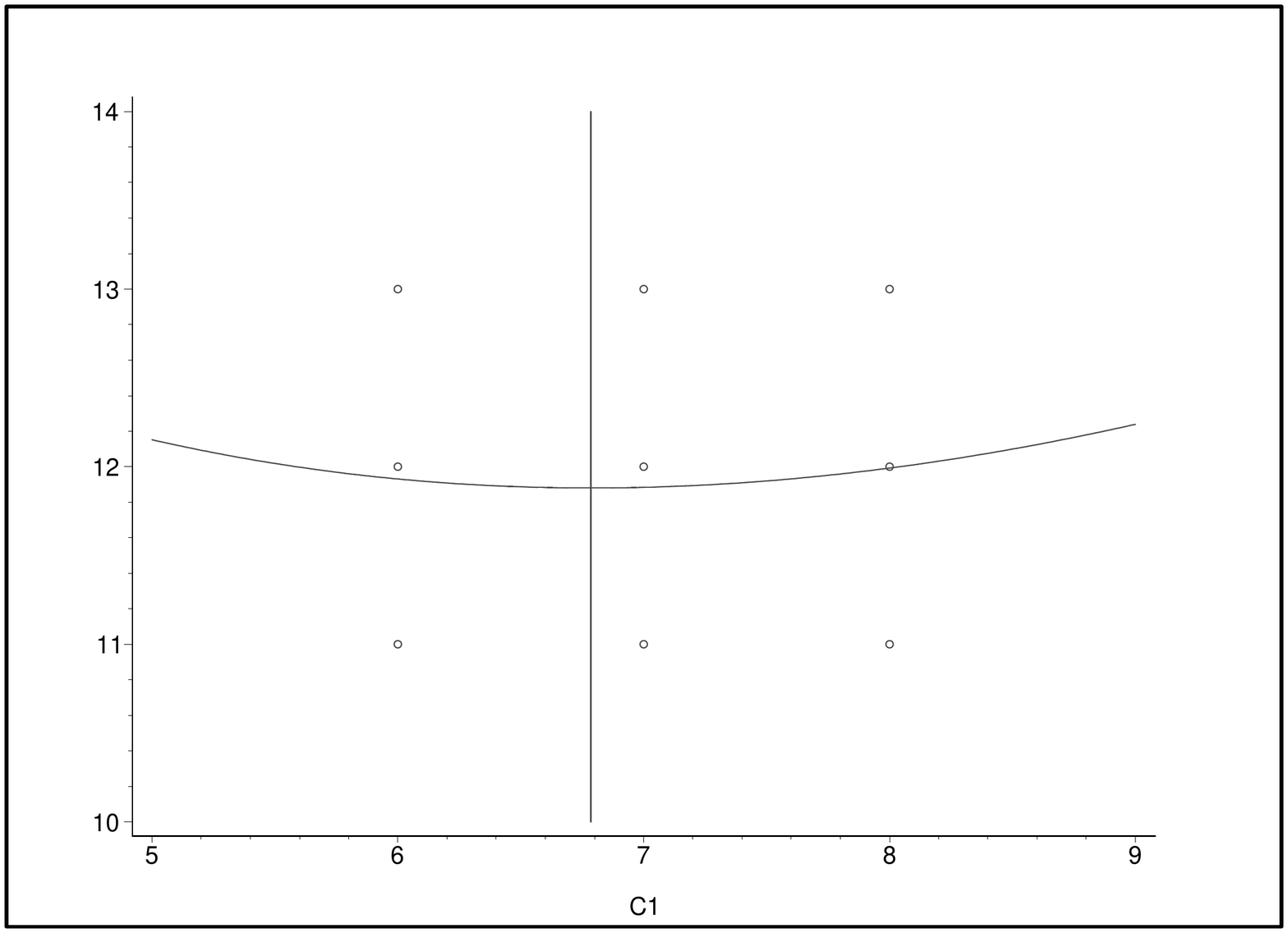}
	\caption{$C^*$  (vertical line), $\FI_2(C)$ (curved line) , $p=0.09,p'=0.05$, and some closest discrete points }
	\label{F2}
\end{figure}

We must now investigate the discrete values, close to $C^*,D^*$, leading to the optimal success probabilities. Of course, it is not the discrete values just  closest to $C^*,D^*$ . We must compute the corresponding numerical values of $w_{C,D}$. For instance, with 
$p=0.09,p'=0.05$,we have $C^*=6.785137352\ldots,D^*=11.88032106\ldots$. The Figure \ref{F2} shows $C^*,\FI_2(C) $ and some closest discrete points. It  appears that, numerically, the discrete solution is 
$C_d^*=7,D_d^*=12$. This fits with the numerical experiments done in \cite{DE15}, with $w_{j,k}, n=40$.
 This gives $ w_{C_d^*,D_d^*}=0.529870739109\ldots$, not far from the continuous value $w_{C^*,D^*}$.

Notice that two discrete couples can lead to the same optimal solution. For instance, with $p'=0.05$,$w_{6,12}-w_{7,12}$ is null
 for $p=0.09396249862111\ldots$.
\subsection{The optimal solution for  $p=p'$}               \label{S23}

Notice that, if $p = p'$, the coefficient of $q^C$ in (\ref{E5}) is null
and the coefficient of $\qt^C$ becomes $T := 2\qt p'^{2} (\ln(q) - \ln(\qt))$. Hence we have the explicit solution

\beq
\mathit{C_{eq}^*} = {\displaystyle \frac {\mathrm{ln}\lp
{\displaystyle \frac {(1 - 2\,p)\,(2\,\mathrm{ln}(1 - p)\,p^{2}
 - 2\,p^{2}\,\mathrm{ln}(1 - 2\,p))}{p^{3}}} \rp}{\mathrm{ln}\lp
{\displaystyle \frac {1 - p}{1 - 2\,p}} \rp}} .                               \label{E70}
\eeq
From (\ref{E6}), we obtain

\[
{\FI_{2,eq}(C)} = \lp - p + p\,\mathrm{ln}(q)\,{C} - 2
\,\mathrm{ln}(q) + 2\,\mathrm{ln}(q)\,p 
\mbox{} + 2\,\mathrm{ln}(q)\,q^{-C}\,(1
 - 2\,p)^{C+1}\rp/(p\,\mathrm{ln}(q)), 
\]
and again, $D_{eq}^*=\FI_{2,eq}(C_{eq}^*)$.
 $w_{C,D},w_{C,C}$ become now
\bal
w_{eq,C,D}&=(D-C)p q^D+q^{D-C}2\lp q^{D+1}-\qt^{C+1}\rp,\non\\
w_{eq,C,C}&=2\lp q^{C+1} -\qt^{C+1}\rp.                                        \label{E7}
\end{align}

\begin{figure}[htbp]
	\centering
		\includegraphics[width=0.8\textwidth,angle=0]{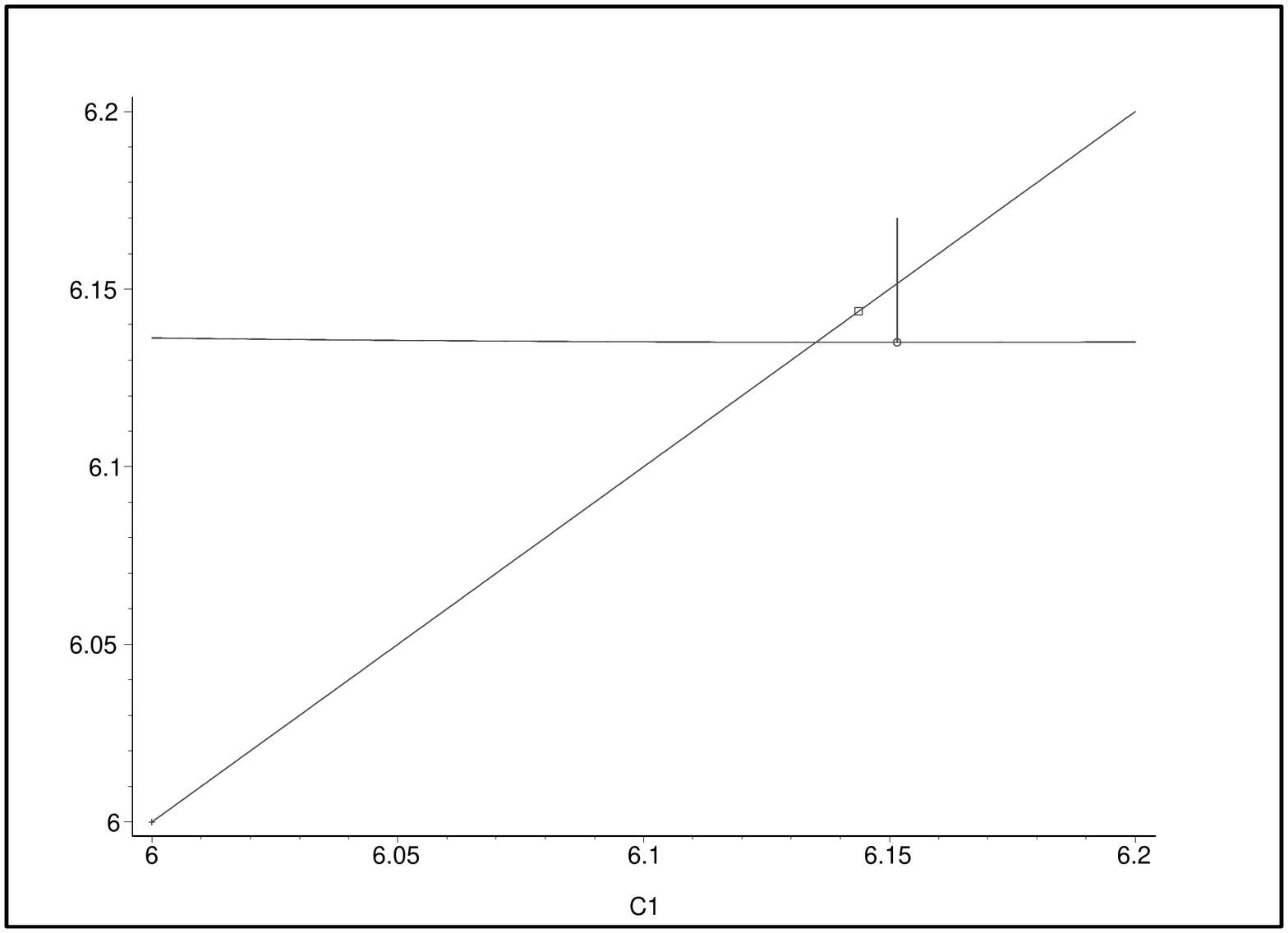}
	\caption{$C_{eq}^*$  (vertical line), $\FI_{2,eq}(C)$ (curved line), $D_{eq}^*$ (circle), $C_{diag}^*$ (square), $(6,6)$ (cross) and the diagonal,
		$p=p'=0.09$ }
	\label{F3}
\end{figure}

Of course, we must use $w_{eq,C,C}$ in our case, and the solution of $\frac{\partial w_{eq,C,C}}{\partial C}=0$ is given by
\[
C_{diag}^*=-(\ln(\ln(q)/\ln(\qt))+\ln(q)-\ln(\qt))/(\ln(q)-\ln(\qt)).
\]
Figure \ref{F3} shows, for $p=p'=0.09 ,C_{eq}^*=6.15156149309\ldots,\FI_{2,eq}(C),D_{eq}^*=6.13502664794\ldots,
C_{diag}^*=6.14370678209\ldots$ the point $(6,6)$ and the diagonal. Notice that the point $(C_{eq}^*,D_{eq}^*)$ is
\textit{below the diagonal}. Of course, only the part $C\leq D$  is relevant. 

 We have $ w_{C_{eq}^*,D_{eq}^*}=0.535056305018\ldots$,  this the maximum, but we can not use it. 
 $ w_{C_{eq}^*,C_{eq}^*}=0.535055655126\ldots$,
$ w_{C_{diag}^*,D_{diag}^*}=0.535055963810\ldots$ is the optimal diagonal continuous value. $ w_{6,6}=0.534951097574\ldots$ is the optimal useful discrete value. We observe the order: $w_{C_{eq}^*,D_{eq}^*}> w_{C_{diag}^*,D_{diag}}^*>w_{C_{eq}^*,C_{eq}^*}>w_{6,6}$.

\begin{figure}[htbp]
	\centering
		\includegraphics[width=0.8\textwidth,angle=0]{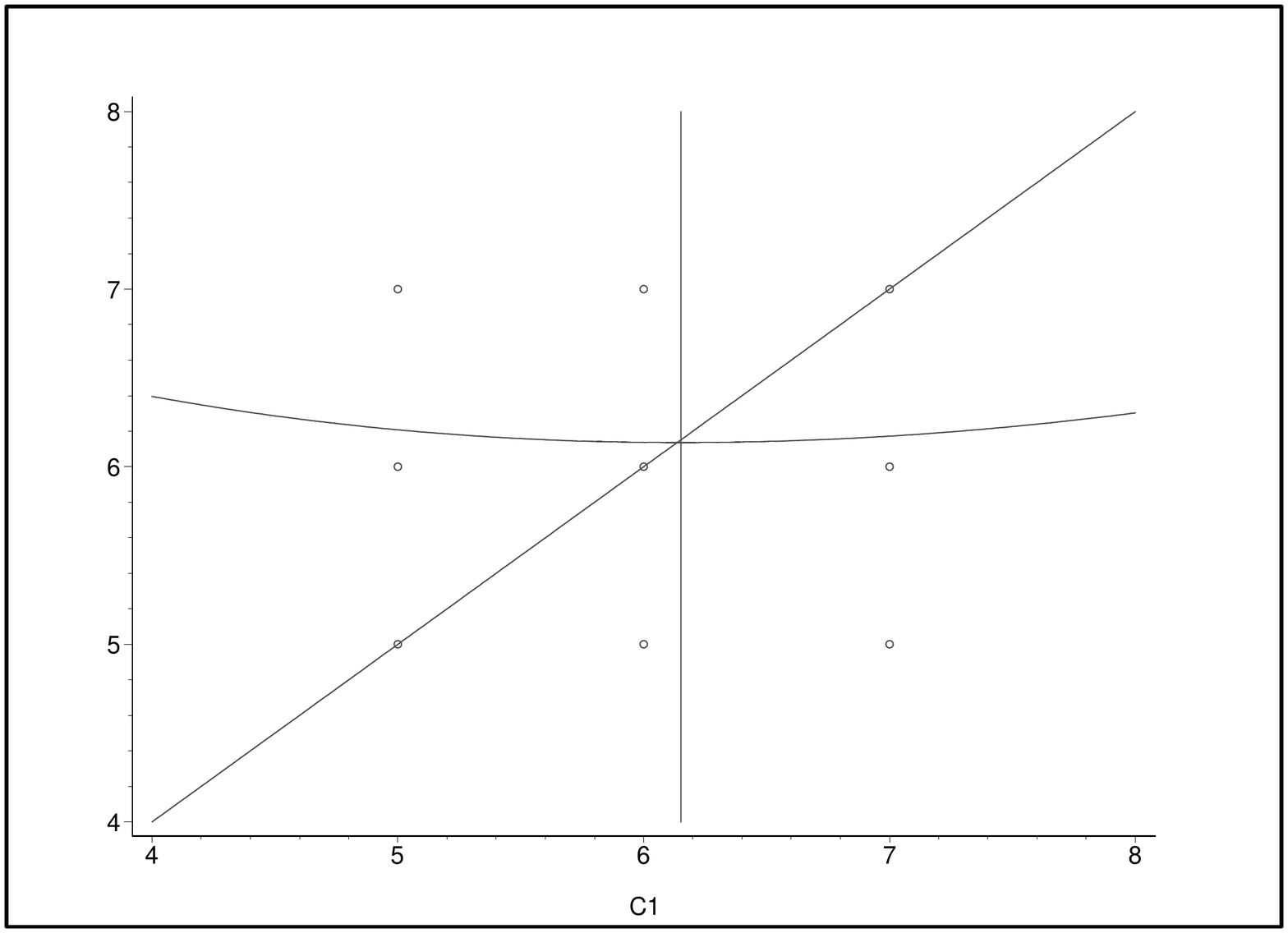}
	\caption{$C^*$  (vertical line), $\FI_2(C)$ (curved line) , $p=0.09,p'=0.08999$, and some closest discrete points  }
	\label{F21}
\end{figure}

\begin{figure}[htbp]
	\centering
		\includegraphics[width=0.8\textwidth,angle=0]{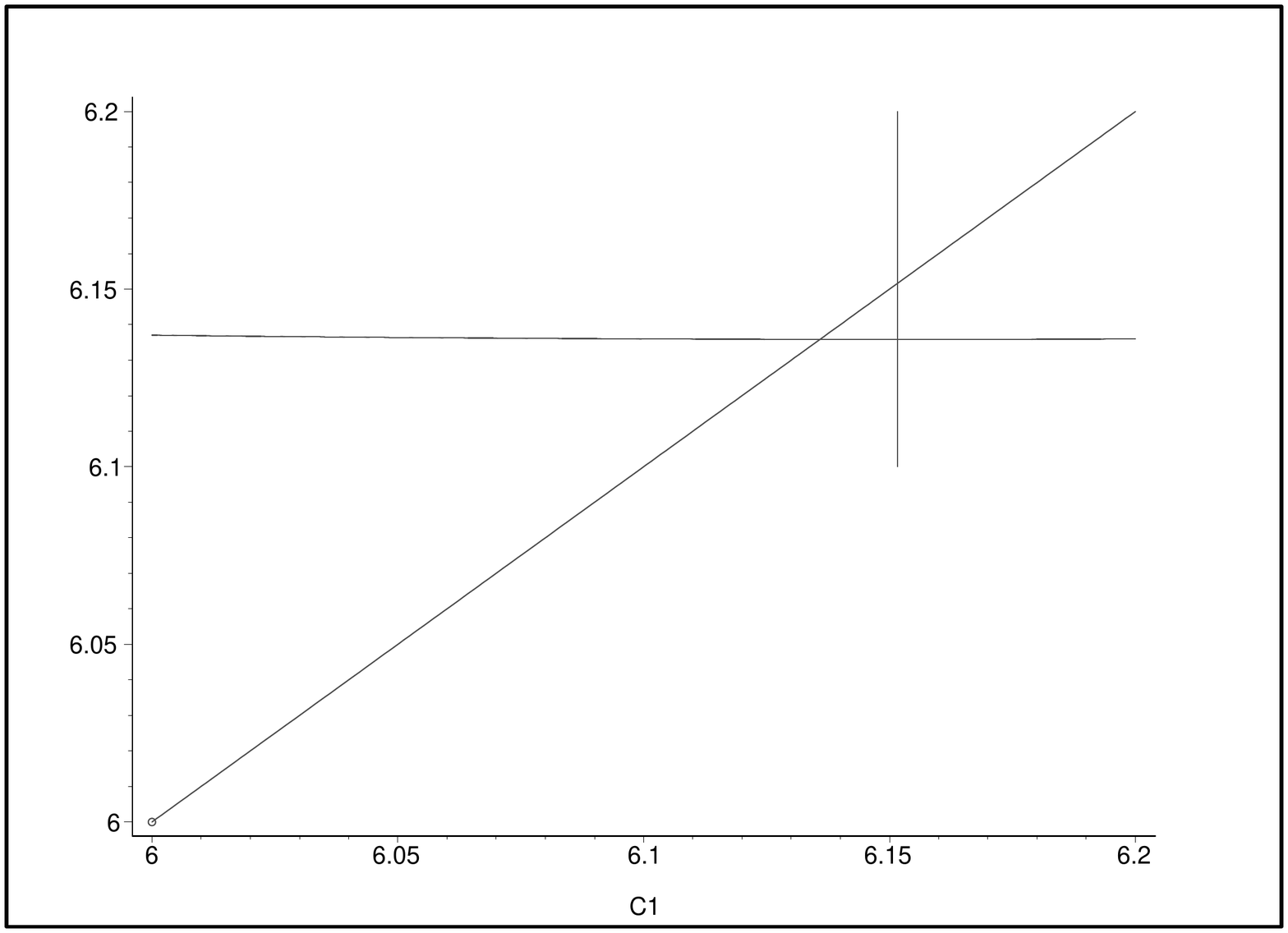}
	\caption{Closer look at Fig. \ref{F21}, with optimal point $(6,6)$ }
	\label{F4}
\end{figure}

 We notice that, even if $p>p'$, we can have a similar situation. If we choose for instance $p=0.09,p'=0.08999$, we have the case described in Figure \ref{F21} and, with a closer look, in  Figure \ref{F4}, where the discrete optimal point $(6,6)$  is on the diagonal. This confirms to the existence of $\gam_3(p)$ defined above.

\begin{figure}[htbp]
	\centering
		\includegraphics[width=0.8\textwidth,angle=0]{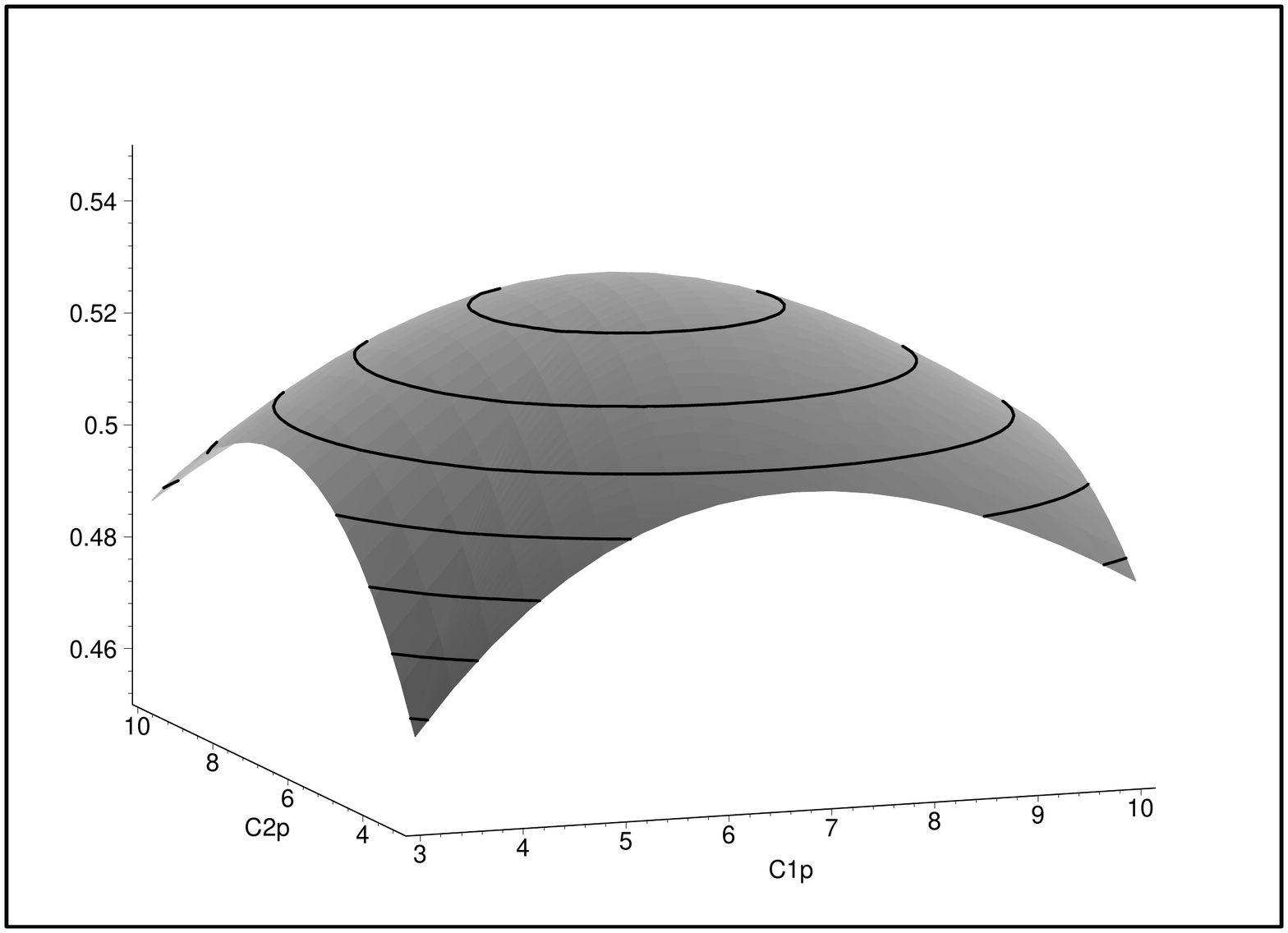}
	\caption{ $w_{C,D},C \leq D $  and $w_{D,C},C \geq D $,$ p=p'=0.09$ }
	\label{F6}
\end{figure}

A plot of $w_{C,D},C \leq D $  and $w_{D,C},C \geq D $,$ p=p'=0.09$ is given in Figure \ref{F6}. This surface is symmetric
 w.r.t.  the diagonal.

\section{The x-strategy}\label{S3}
We recall the notion of an x-strategy given in the Introduction:   let $U_i,i=1,2,\ldots,n$  be independent random variables uniformly distributed on the  interval $[0,1]$. Let $T_i=U_{\{i\}}$: $T_i$ is the $i$th order statistic of the $U_i$'s. $T_i$ is the arrival time of $X_i$. The strategy is to wait until some time $x_n^*$ and from $x_n^*$ on, we select the 
first $X_i=+1$ or   $X_i=-1$, using the previous algorithm with $p=p'$. Following Bruss \cite{Br841}, we call this strategy an x-strategy. In \cite{DE15}, the author gives,  for this problem, the optimal $x_n^*$ and  the corresponding success probability $P_n^*$.
In this Section, we analyze accordingly  asymptotic expansions  for $p=p'$ . We also consider the  success probability for small $p$,  and also the case $p>p'$.
\subsection{The x-strategy, $p=p'$}

Let first recall a few results from \cite{DE15}. If we denote by $\ell$ the number of observed  variables, starting from $x$, we must set, in (\ref{E7}), $C=\ell-1$. This leads to the success probability
\[
{P_n(x,p)} = \sum_0^n \bin{n}{ \ell}(1-x)^\ell x^{n-\ell } 2\lp q^{\ell} -\qt^{\ell}\rp
= 2\,\lp(q + p\,x)^{n} - \,(2\,q - 1 + 2\,p \,x)^{n}\rp.
\]
The optimal value $x_n^*$ is solution of $\frac{d P_n(x,p)}{dx}=0$, which leads to
\[
{x_n^*} := {\displaystyle \frac { - q + 2\,{\be_n}\,q - 
{be}}{1 - q - 2\,{\be_n} + 2\,{be}\,q}} ,\B \be_n:=2^{1/(n-1)}.
\]
This gives

\[
{P_n^*:=P_n(x_n^*,p)} := 2\,(2\,2^{(\frac {1}{{n} - 1})
} - 1)^{(1 - {n})}.
\]
Notice that $P_n^*$ \textit{is independent of $p$}. Open Problem $1$: why is it so?
It appears that, for $p=\tilde{p}_n$, we have $x_n^*=0$, with
\[\tilde{p}_n=\frac{\be_n-1}{2\be_n-1}.\]
We can also check that $P_n(0,\tilde{p}_n)=P_n^*$.

Let us now turn to the the  asymptotic analysis of the case $p=p'$ and the corresponding behaviour for small $p$.

 Asymptotically, we obtain, for $n\ra \infty$,
\beq
x_n^* = 1-\frac{\ln(2)}{np}+  {\displaystyle \frac {1}{2}} \,{\displaystyle 
\frac {\mathrm{
ln}(2)\,( - 2 + 3\,\mathrm{ln}(2))}{ pn^2}}+\BO\lp \frac{1}{n^3}\rp, \label{E81}
\eeq
\[
P_n^*={\displaystyle \frac {1}{2}}  + 
{\displaystyle \frac {1}{2}} \,{\displaystyle \frac {\mathrm{ln}(
2)^{2}}{{n}}}  + \frac{{\displaystyle \frac {1}{4}} \,\mathrm{ln}(2)^{2}\,(2 -
 2\,
\mathrm{ln}(2) + \mathrm{ln}(2)^{2})}{n^2}+\BO\lp \frac{1}{n^3}\rp,                                         
\]
\beq
\tilde{p}_n= {\displaystyle \frac {\mathrm{ln}(2)}{{n
}}}  + {\displaystyle \frac {  - {\displaystyle \frac {1}{2}} \,\mathrm{ln}(2)
\,
( - 2 + 3\,\mathrm{ln}(2))}{{n
}^{2}}} +\BO\lp \frac{1}{n^3}\rp  .                                                                \label{E8}
\eeq
$P_n^*$ converges to $1/2$ for $n\ra \infty$ .For instance, $P_{500}^*=0.500480981417\ldots$.
An interesting question is: what is the behaviour of $P_n^*$ for $p\leq \tilde
{p}_n$? Following (\ref{E8}), we tentatively set $q=1-y/n,x=0$ in $P_n(x,p)$. This leads to
\[P_n(y)= 2\,e^{-y} - 2\,e^{-2y} + 
{\displaystyle \frac { - e^{-y}\,y^{2} + 4\,e^{-2y}\,y
^{2}}{{n}}}  + {\displaystyle \frac {2\,e^{ - y}\,( - {\displaystyle \frac {1}{3}} \,y^{3} + 
{\displaystyle \frac {1}{8}} \,y^{4}) - 2\,e^{-2y}\,\lp -
{\displaystyle \frac {8}{3}} \,y^{3} + 2\,y^{4}\rp}{{n
}^{2}}}+\BO\lp \frac{1}{n^3}\rp. \]
In order to check, we put the first term of $\tilde{p}_n$ i.e. $y=\ln(2)$ into $P_n(y)$. Expanding, this leads to the first two terms of 
$P_n^*$. Similarly, putting the first two terms of $\tilde{p}_n$, i.e. $y=\ln(2)  + {\displaystyle \frac {  - {\displaystyle \frac {1}{2}} \,\mathrm{ln}(2)
\,
( - 2 + 3\,\mathrm{ln}(2))}{{n}}}$ into $P_n(y)$ gives the  first three terms of $P_n^*$.

\subsection{The x-strategy for  $p>p'$}
This case was not considered before.
We can still use the x-strategy, but now we must set $D=\ell-1$. Also, if $D\geq C_d^* $, we use $w_{C_d^*,D}$ and if 
$D \leq C_d^* $, we use $w_{D,D}$ (we must stay above the diagonal). This leads to
\bals
P_n^*&= \sum_{\ell=C_d^*}^n \bin{n}{ \ell}(1-x)^\ell x^{n-\ell }w_{C_d^*,\ell-1}+
\sum_{l=0}^{C_d^*} \bin{n}{ \ell}(1-x)^\ell x^{n-\ell }w_{\ell-1,\ell-1}\\
&= \sum_0^n \bin{n}{ \ell}(1-x)^\ell x^{n-\ell }w_{C_d^*,\ell-1}
+\sum_{l=0}^{C_d^*} \bin{n}{ \ell}(1-x)^\ell x^{n-\ell }[w_{\ell-1,\ell-1}-w_{C_d^*,\ell-1}].\\
\end{align*}
The first summation leads to $S_1+S_2$, with

\bals
{S_1}& := {\displaystyle \frac {\lp {\displaystyle \frac {(1 - 
x)\,q'}{x}}  + 1\rp^{{n}}\,x^{{n}}\,( - {C_d^*} - 1)\,p'}{q'} }\\
& + 
{\displaystyle \frac {\lp{\displaystyle \frac {(1 - x)\,q'}{x}}  + 1\rp^{{n}}\,(1 - x)\,q'\,{n}\,\lp
 - {\displaystyle \frac {x^{{n}}\,{C_d^*
}\,p'}{q'}}  - 
{\displaystyle \frac {x^{{n}}\,( - {
C_d^*} - 1)\,p'}{q'}} \rp}{x\,\lp{\displaystyle \frac {(1 - x)\,(1 - {p'
})}{x}}  + 1\rp}} ,
\end{align*}

\[
{S_2} := \lp{\displaystyle \frac {(1 - x)\,q'}{x}}  + 1\rp^{{n}}\,x^{{n}}\,q'^{ - {C_d^*} - 1}\lp 
{\displaystyle \frac {{p}\,(q^{{C_d^*} + 1} - \tilde{q}^{{C_d^*} + 1})}{{p'}}} 
 \mbox{} + {\displaystyle \frac {p'\,(q'^{{C_d^*} + 1} - \tilde{q}^{{C_d^*} + 1})}{{p}}} \rp.
\]
The second summation leads to a complicated expression, involving binomials and hypergeometric terms that we do not display here. However, if we plug in numerical values, for instance $p=0.09,p'=0.05,n=40, C_d^*=7$, we obtain a tractable function $P(x)$ that we can differantiate, leading to $x^*=0.667967251301\ldots$.  This gives $P(x^*)=0.523618813813\ldots$.
\newpage
\section{The x-strategy with incomplete information }\label{S4}
Bruss suggested to analyze this strategy because incomplete information has an increased appeal for applications.

We will only consider the case $p=p'$. The other cases can  similarly analyzed, with more complicated algebra. We will consider the cases $p$ known, $n$ unknown, then $n$ known, $p$ unknown and finally $n,p$  unknown. Some simulations are also provided. In all our numerical expressions, we will use  $n=500,p=0.03$.
\subsection{The case $p$ known, $n$ unknown}
We will always denote by $m$ the number of observed variables up to time $x$ and by $k$ the number of $\{+1,-1\}$ observed variables up to time $x$.  From (\ref{E81}), we have 
$x_n^*\sim 1-\frac{\ln(2)}{np}$ and we will use the natural estimate $\tilde{n}=\frac{m}{x}$. Hence we start from the formal 
equation resulting from (\ref{E81}), hence
\[x=1-\frac{x\ln(2)}{mp},\]
from which we deduce the two functions
\bals
x&=g(m,p)=\frac{mp}{mp+\ln(2)},\\
m&=f(x,p)=\frac{\ln(2)x}{p(1-x)}.
\end{align*}

\begin{figure}[htbp]
	\centering
		\includegraphics[width=0.8\textwidth,angle=0]{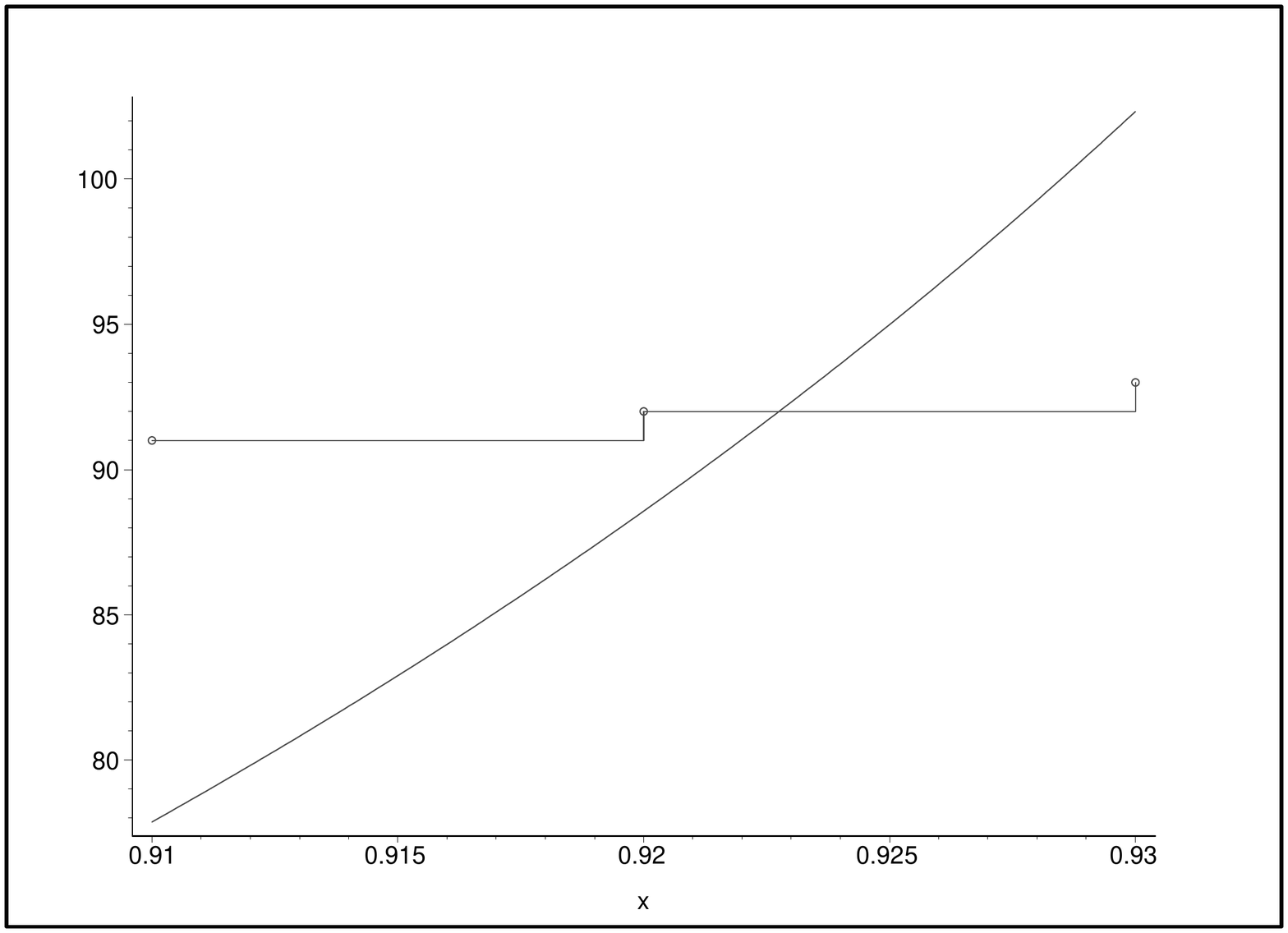}
	\caption{The case $p$ known, $n$ unknown: a typical  crossing because it occurs close to $1$ }
	\label{F7}
\end{figure}

\begin{figure}[htbp]
	\centering
		\includegraphics[width=0.8\textwidth,angle=0]{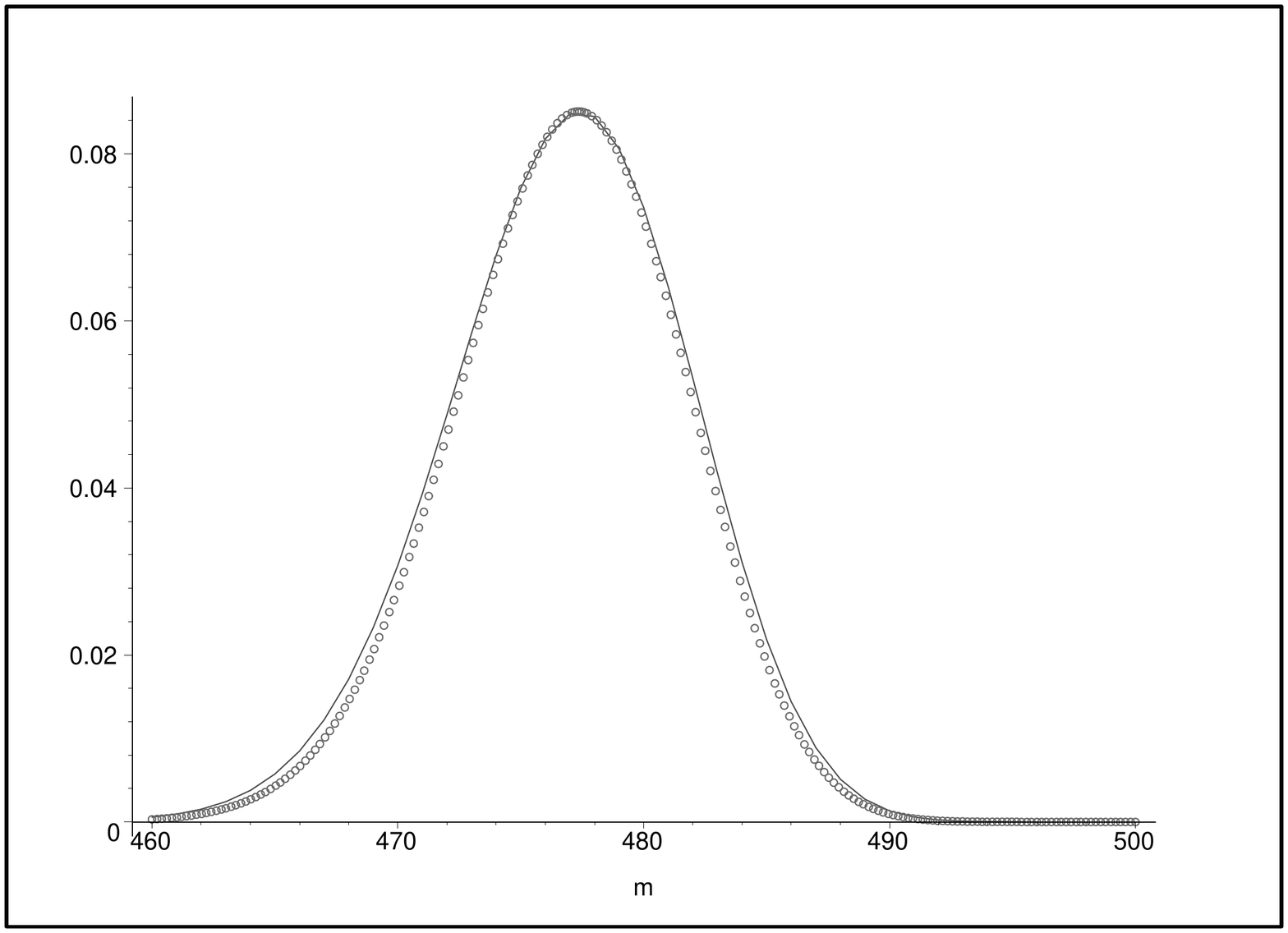}
	\caption{The case $p$ known, $n$ unknown: $\Fi(n,\mu,p)$ (line) , with $G(n,\mu,x_n^*) $ (circles) }
	\label{F8}
\end{figure}

Our algorithm proceeds as follows: wait until $m$ crosses the function $f(x,p)$ at value $m^*$.  It follows from  Bruss and Yor \cite{BY12} ,Thm 5.1  that all optimal actions are confined to the interval $[x_1,1]$ for some $x_1<1$ so that we can ignore preceding crossing, if any. (In the last-arrival problem, supposing no information at all, this value $x_1$ equals $1/2$). The crossing algorithm  gives  a value    $x^*=g(m^*,p)$. We will use this value  in the x-strategy. First of all we notice that, asymptotically, $m$ corresponds to a Brownian bridge of order $\sqrt{n}$ with a drift $nx$. On the other side, $f'(x_n^*) \sim pn^2/\ln(2)$. Hence, with high probability, $m$ crosses $f(x,p)$ only  once in the neighbourhood of $x_n^*$.  Let
\[G(n,m,x):=\bin{n}{m}x^m(1-x)^{n-m}\]
be the distribution of $m$ at time $x$. We have
\[\Fi(n,\mu,p):=\P(m^*=\mu)\sim G(n,\mu,g(\mu,p)),\]
and using 
\[P_{eq}(\ell,p):=2[q^\ell-\qt^{\ell} ],\]
we obtain the success probability
\[P(n,p)= \sum_1^n \Fi(n,\mu,p)P_{eq}(n-\mu,p).\]
For instance,  we show in Figure \ref{F7} an illustration of a typical  crossing and in Figure \ref{F8}, 
the function $\Fi(n,\mu,p)$ (line) together with $G(n,\mu,x_n^*) $ (circles) (the classical  x-strategy $\mu$ distribution ).

 The distributions are quite similar. Open Problem $2$: why? We obtain $P(n,p)\sim  0.5234\ldots$  (In the numerical summations, we sum $\mu$ from some value $\tilde{\mu}$ to avoid any problems near the origin)

\subsection{The case $n$ known, $p$ unknown}
Now we use the following estimate for $p:\tilde{p}=k/(2m)$. The formal starting equation is
\[x= 1-\frac{\ln(2)}{np}.\]
Hence the two functions
\bals
x&=u(n,p)=1-\frac{\ln(2)}{np},\\
p&= h(n,x)=\frac{\ln(2)}{n(1-x)}.
\end{align*}
The algorithm waits until $\tilde{p}$ crosses  function $ h(n,x)$ at value $p^*$, giving a value $x^*=u(n,p^*)$. Again, with high probability, $\tilde{p}$ crosses $ h(n,x)$ only  once in the neighbourhood of $x_n^*$. The joint distribution of $m,k$ at time $x$  is given, with $k\leq m$ by
\[H(n,m,k,x,p)=G(n,m,x)\bin{m}{k}(2p)^k (1-2p)^{m-k}.\]
The joint distribution of $m=\mu,k$ \textit{given} that $\tilde{p}$ has just crossed $ h(n,x)$ is given by   
\[\Pi(n,\mu,k,p)\sim H(n,\mu,k,u(n,\tilde{p}),p).\]
We have
\[\Fi(n,\mu,p):=\P(m^*=\mu)\sim \sum_{k=1}^\mu \Pi(n,\mu,k,p), \]
and finally the success probability is given by
\[P(n,p)= \sum_1^n \Fi(n,\mu,p)P_{eq}(n-\mu,p).\]

\begin{figure}[htbp]
	\centering
		\includegraphics[width=0.8\textwidth,angle=0]{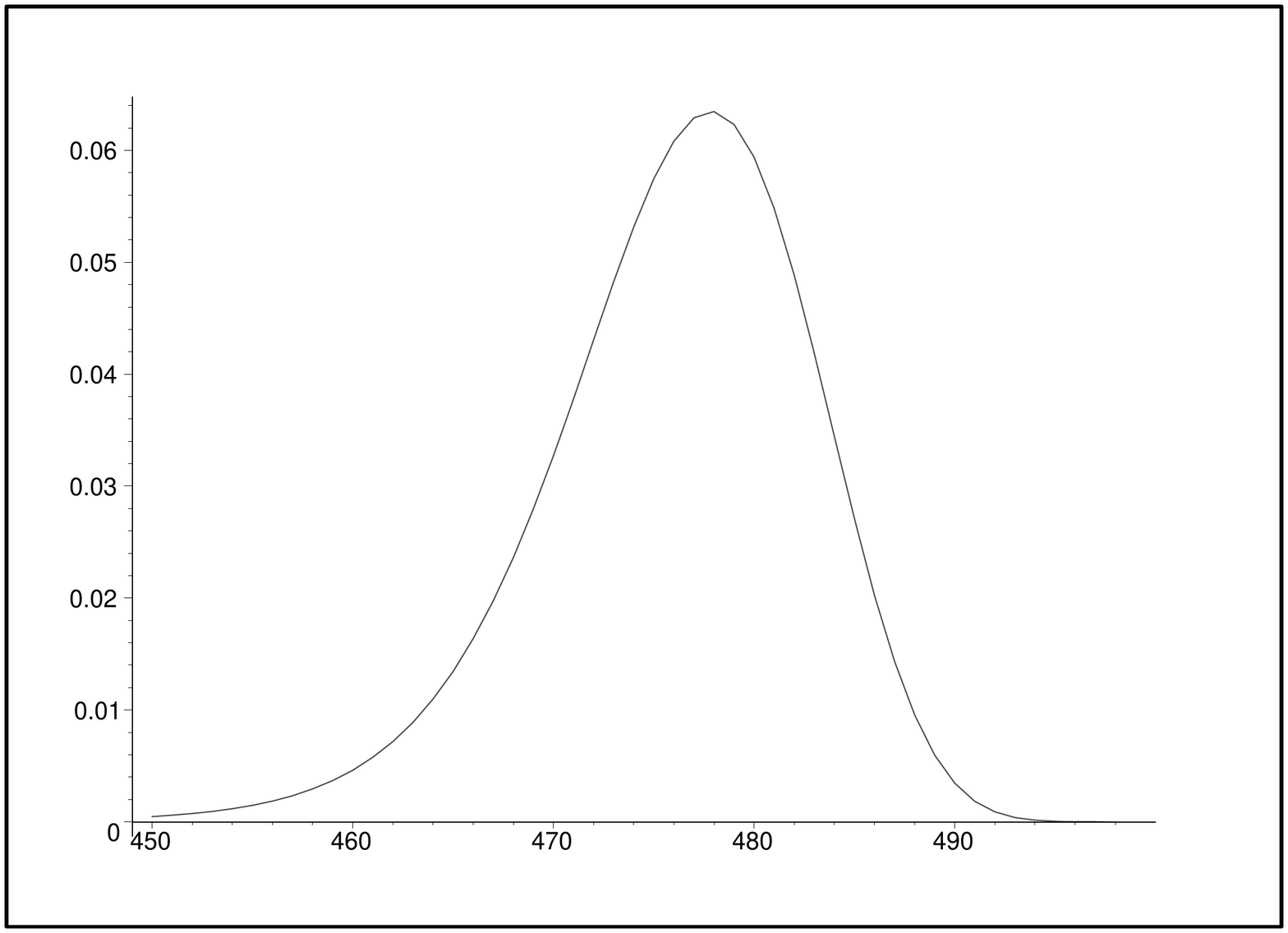}
	\caption{The case $n$ known, $p$ unknown: $\Fi(n,\mu,p)$ }
	\label{F9}
\end{figure}

As an example,  we show in Figure \ref{F9} 
the function $\Fi(n,\mu,p)$. Also $P(n,p)\sim  0.4927\ldots$

\subsection{The case $n,p$ unknown}
The estimates are now $\tilde{p}=k/(2m),\tilde{n}=\frac{m}{x}$. This leads to formal starting equation
\[x=1-\frac{2x\ln(2)}{k}.\]
Hence the two functions
\bals
x&=v(k)=\frac{k}{k+2\ln(2)},\\
k&= w(x)=\frac{2\ln(2)x}{(1-x)}.
\end{align*}
The algorithm waits until $k$ crosses  function $ w(x)$ at value $k^*$, giving a value $x^*=v(k^*)$.
Again, with high probability, $k$ crosses $ w(x)$ only  once in the neighbourhood of $x_n^*$.
The joint distribution of $m=\mu,k$ \textit{given} that $k$ has just crossed $ w(x)$ is given by   
\[\Pi(n,\mu,k,p)\sim H(n,\mu,k,v(k),p).\]
We have
\[\Fi(n,\mu,p):=\P(m^*=\mu)\sim \sum_{k=1}^\mu \Pi(n,\mu,k,p), \]
and finally the success probability is given by
\[P(n,p)= \sum_1^n \Fi(n,\mu,p)P_{eq}(n-\mu,p).\]

\begin{figure}[htbp]
	\centering
		\includegraphics[width=0.8\textwidth,angle=0]{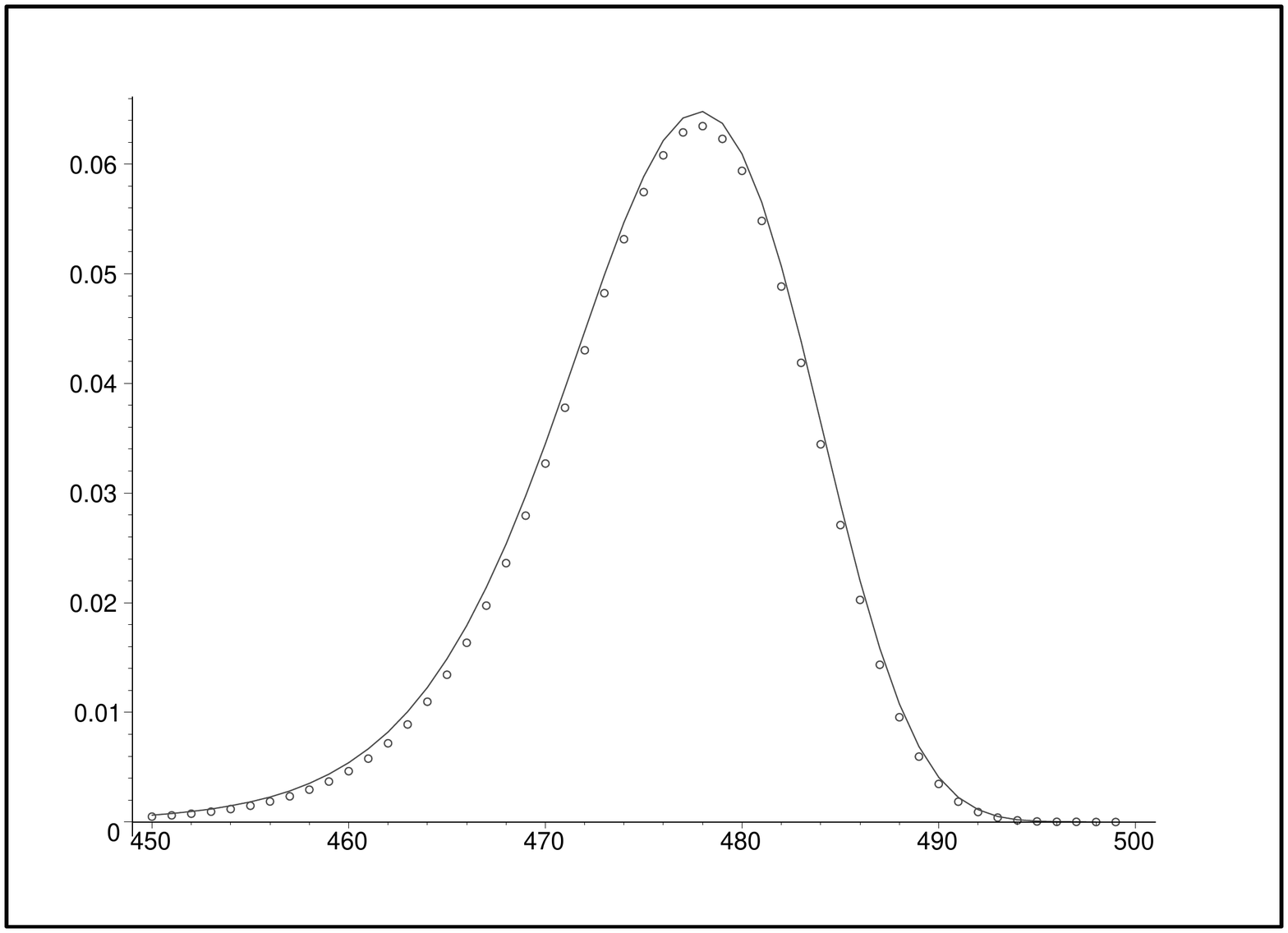}
	\caption{The case $n,p$ unknown: $\Fi(n,\mu,p) $ (line) together with the corresponding distribution in the the case $n$ known, $p$ unknown (circles)}
	\label{F11}
\end{figure}

For instance, we show in Figure \ref{F11} 
the function $\Fi(n,\mu,p) $ together with the corresponding distribution in the the case $n$ known, $p$ unknown (circles).
Curiously enough, the distributions are quite similar but different from the case  $p$ known, $n$ unknown. Open Problem $3$: why?
Also $P(n,p)\sim  0.5156\ldots$.

\subsection{Simulations}
\begin{figure}[htbp]
	\centering
		\includegraphics[width=0.8\textwidth,angle=0]{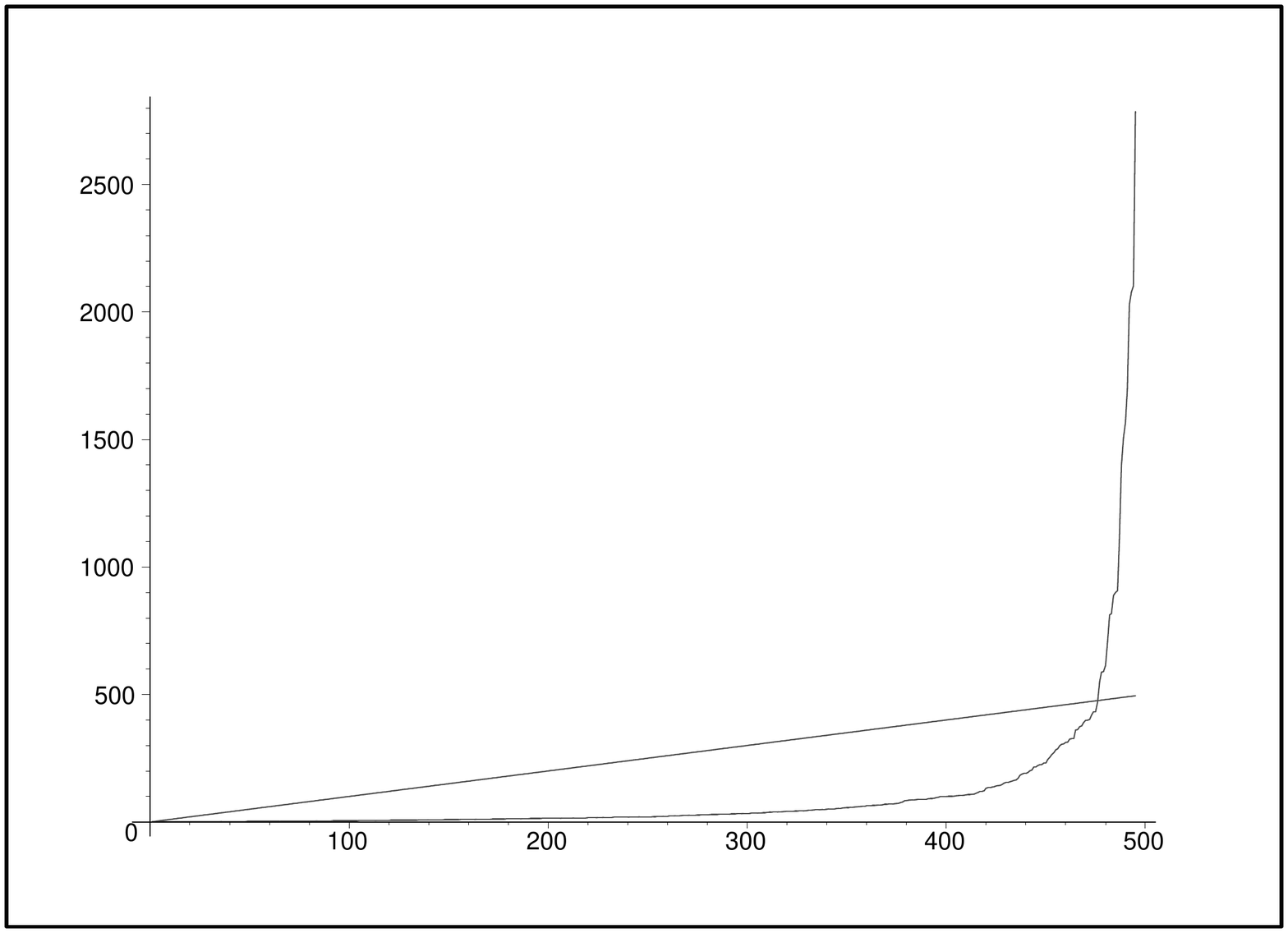}
	\caption{The case $p$ known, $n$ unknown: a typical path }
	\label{F12}
\end{figure}

\begin{figure}[htbp]
	\centering
		\includegraphics[width=0.8\textwidth,angle=0]{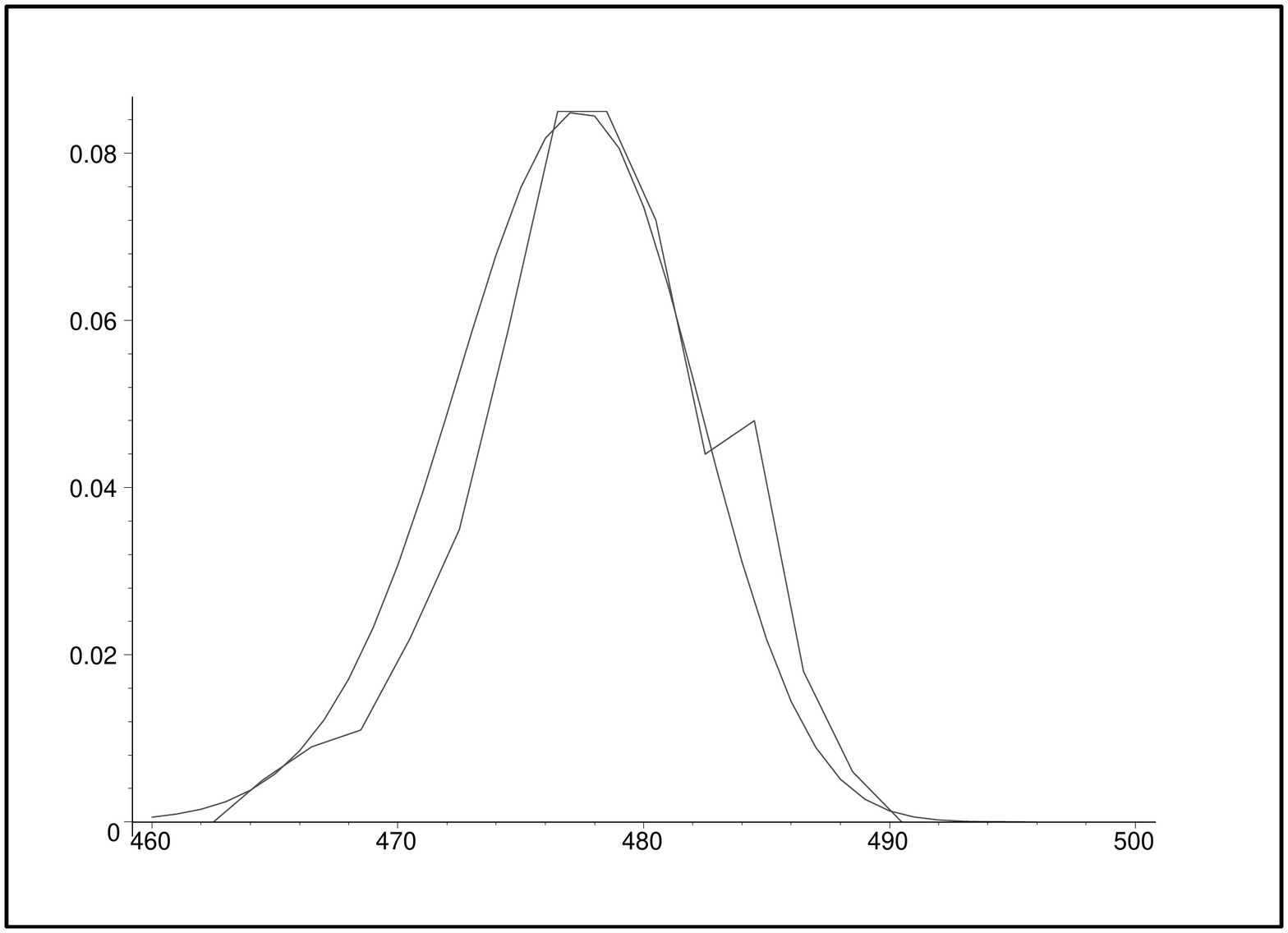}
	\caption{The case $p$ known, $n$ unknown: the empirical observed distribution, together with  $\Fi(n,\mu,p)$ }
	\label{F13}
\end{figure}

We have made three simulations of the crossing value $\mu$ distribution compared with $\Fi(n,\mu,p)$. Each time we made $500$
simulated paths. For the case  $p$ known, $n$ unknown, a typical path is given in Figure \ref{F12} and, in Figure \ref{F13} , we show the empirical observed distribution, together with  $\Fi(n,\mu,p)$ ( For the purpose of smoothing, we have grouped two successive observed probabilities together). Numerically, this gives $P_{sim}(n,p)=0.4981\ldots$.

\begin{figure}[htbp]
	\centering
		\includegraphics[width=0.8\textwidth,angle=0]{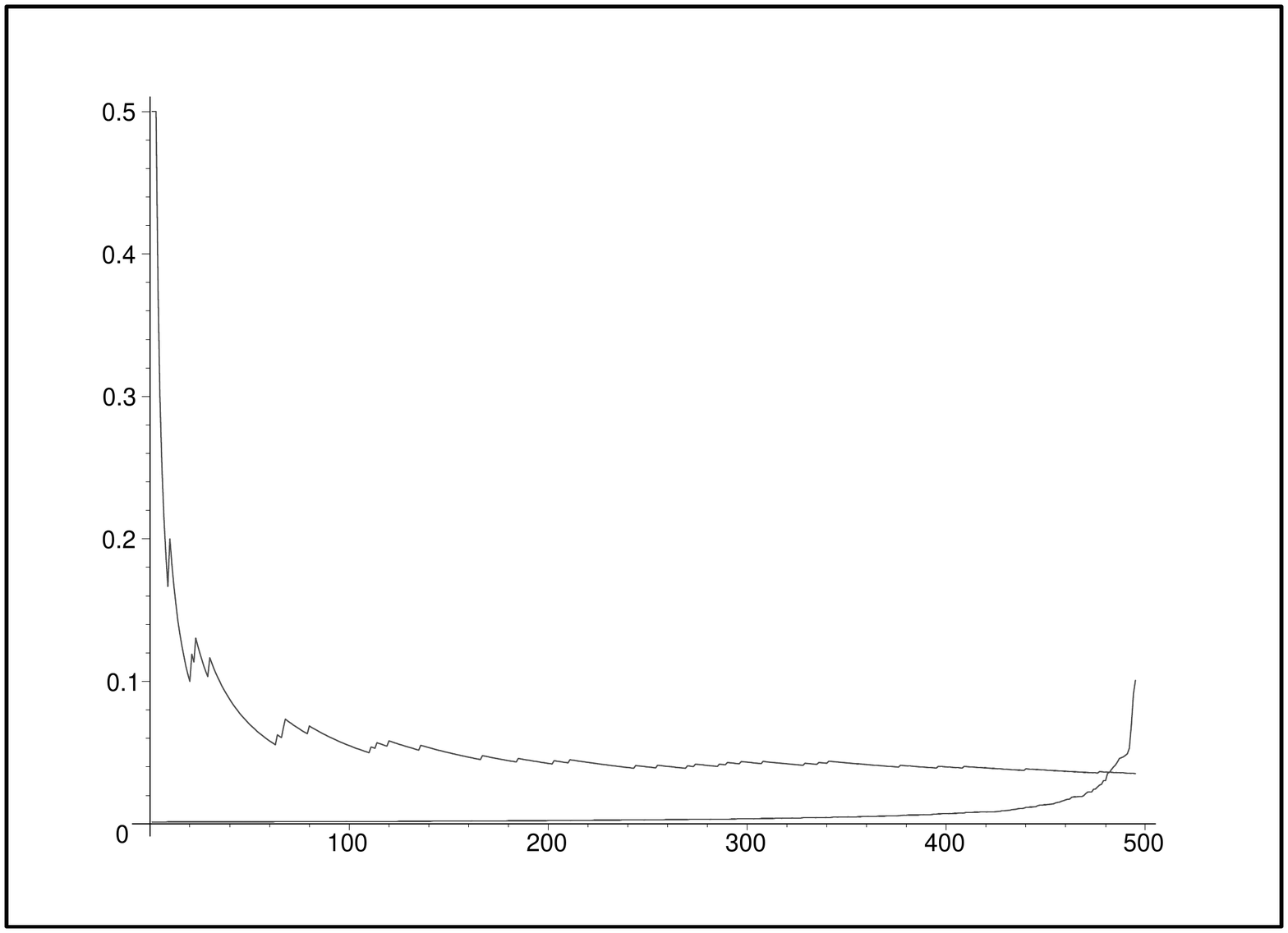}
	\caption{The case $n$ known, $p$ unknown: a typical path }
	\label{F14}
\end{figure}

\begin{figure}[htbp]
	\centering
		\includegraphics[width=0.8\textwidth,angle=0]{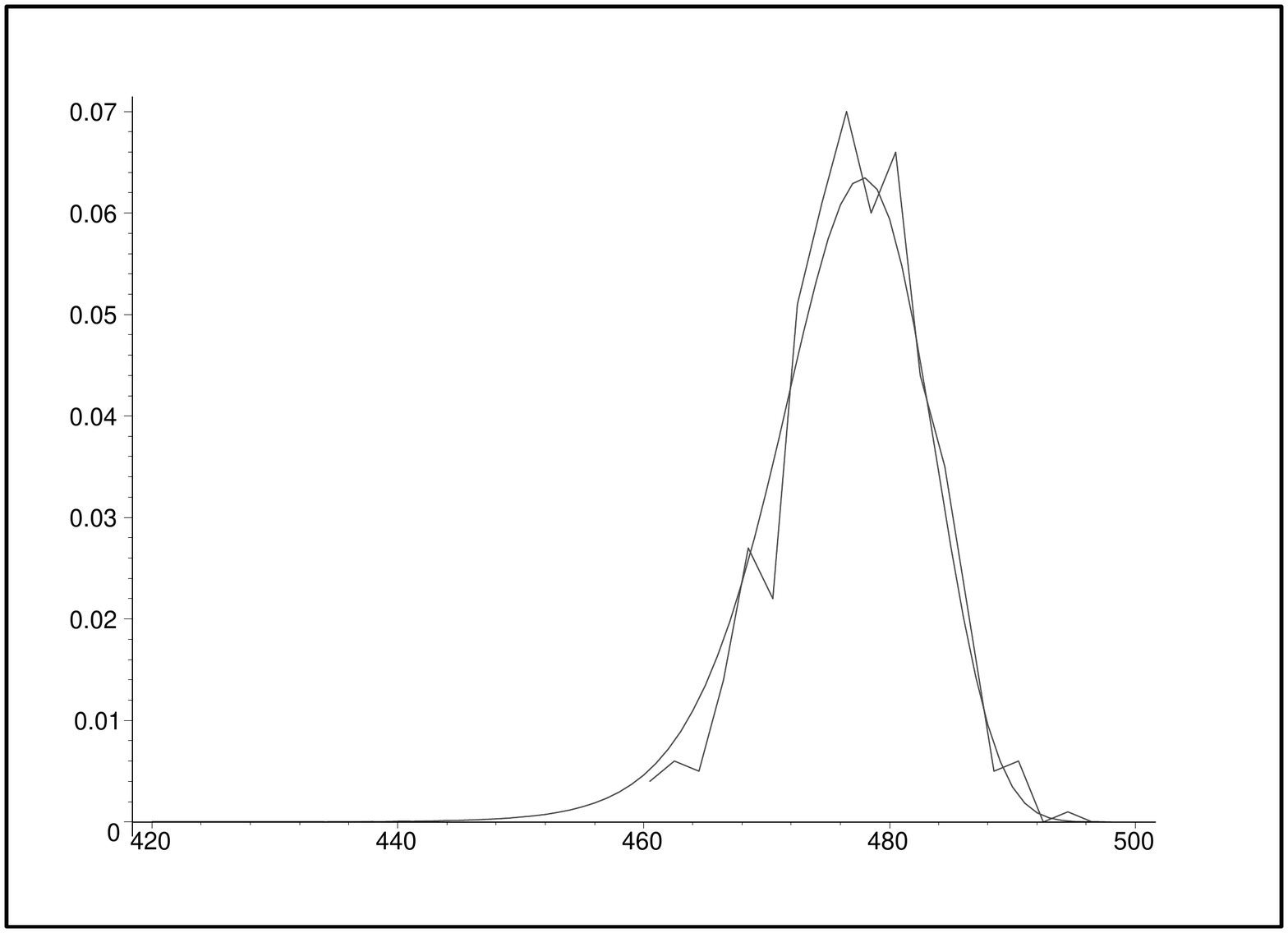}
	\caption{The case $n$ known, $p$ unknown:  the empirical observed distribution, together with  $\Fi(n,\mu,p)$ }
	\label{F15}
\end{figure}

Similarly, for  the case $n$ known, $p$ unknown, a typical path is given in Figure \ref{F14} and, in Figure \ref{F15} , we show the empirical observed distribution, together with  $\Fi(n,\mu,p)$. Numerically, this gives $P_{sim}(n,p)=0.4915\ldots$. 

\begin{figure}[htbp]
	\centering
		\includegraphics[width=0.8\textwidth,angle=0]{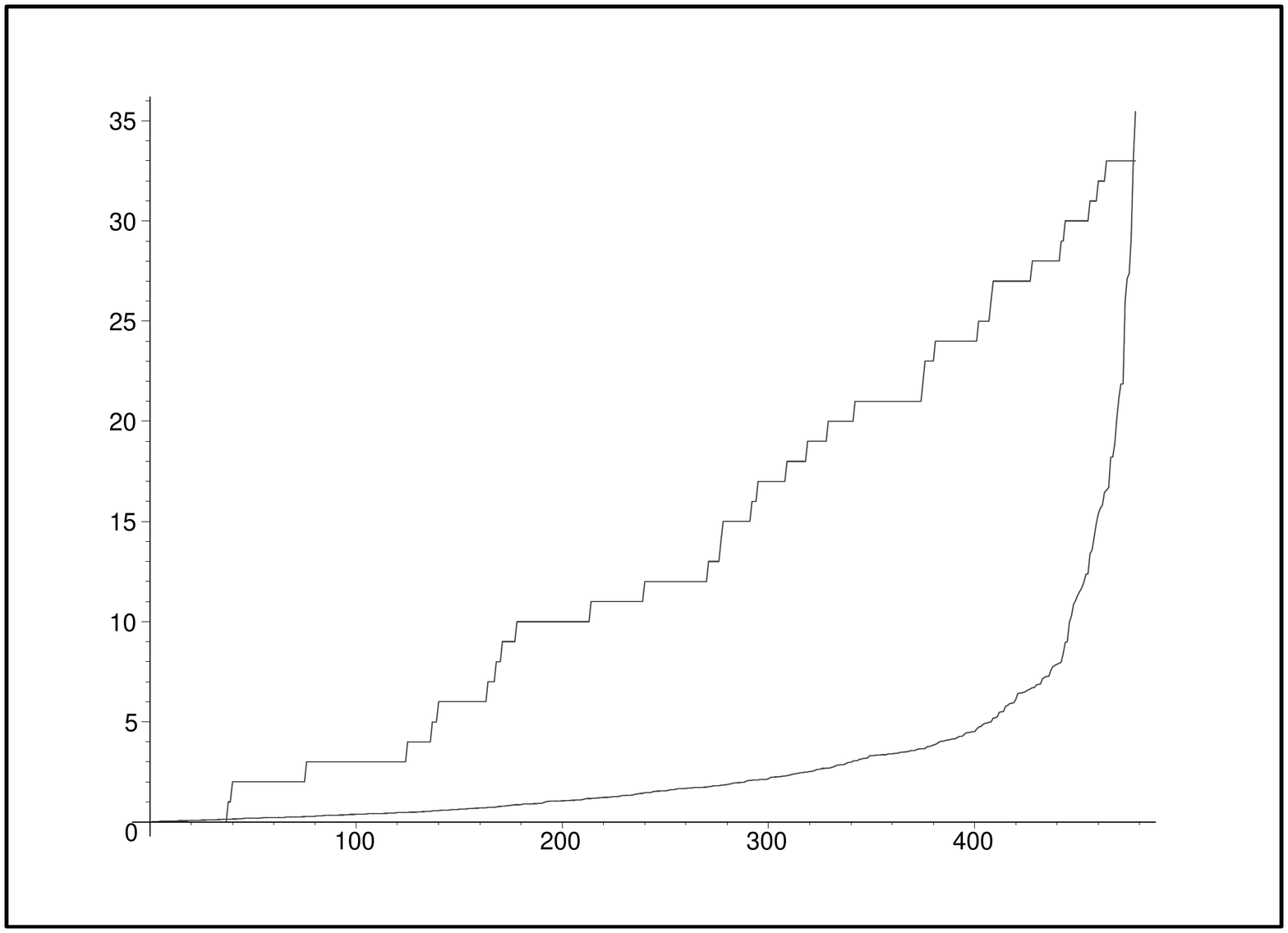}
	\caption{The case $n,p$ unknown:  a typical path  }
	\label{F16}
\end{figure}

\begin{figure}[htbp]
	\centering
		\includegraphics[width=0.8\textwidth,angle=0]{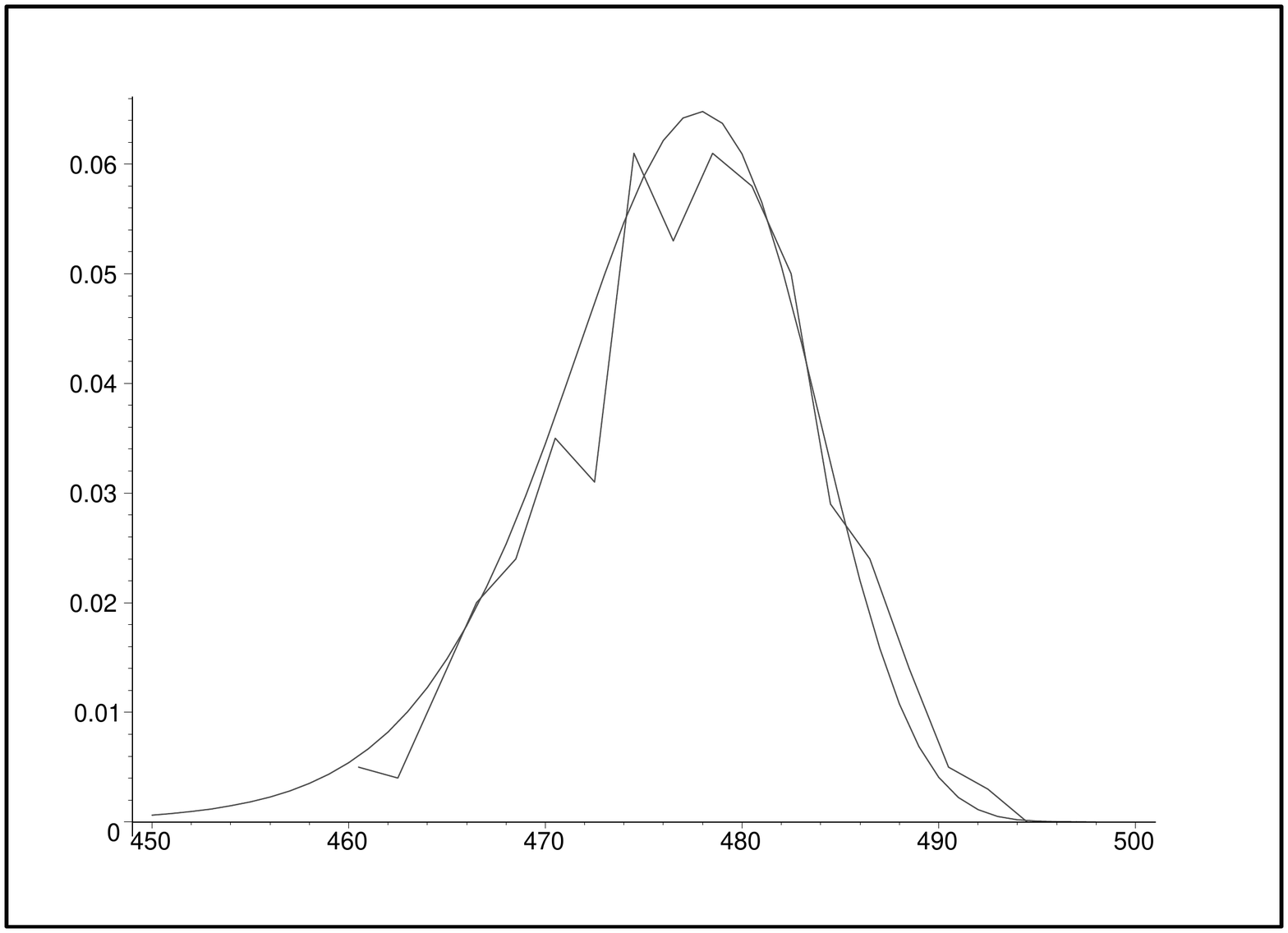}
	\caption{The case $n,p$ unknown: the empirical observed distribution, together with  $\Fi(n,\mu,p)$  }
	\label{F17}
\end{figure}

For the case $n,p$ unknown,  a typical path is given in Figure \ref{F16} and, in Figure \ref{F17} , we show the empirical observed distribution, together with  $\Fi(n,\mu,p)$. Numerically, this gives $P_{sim}(n,p)=0.4805\ldots$.

All fits are  satisfactory. 
\newpage
\section{Conclusion }\label{S5}
Using a continuous model, some asymptotic expansions and an incomplete information strategy, we have obtained a refined and asymptotic analysis of the extended Weber problem and several versions of Bruss-Weber problems. Three problems remain open: 
 why is $P_n^*$  independent of $p$? Can we justify the similarities in the distributions of the crossing value $m^*$? An interesting problem  would be to consider the case with several values
 $\{-k,-(k-1),\ldots,-1,0,1,\ldots,k\}$ with corresponding stopping times. If moreover values can be associated with relative ranks, such problems (Bruss calls them\\ `` basket " problems ) are partially studied in Dendievel \cite{DE16}.

\section{Appendix. An asymptotic analysis of $\gam_4(p)$}                        \label{S6}
Some numerical experiments show that, for $\Cs$ near $-1$,  $\gam_4(p)$ is very close to $p'=1-p$, and that no value $\Cs<-1$ appears as solution of (\ref{E50}). The asymptotic behaviour of $\gam_4(p)$ for $\Cs$ near $-1$ can be summarized as follows.  We keep only dominant terms in our expansions.
\bit
\item  for $p$ near $1$, we set $p'= w$. For $w=0$, $ \FI_1(C)$ is identically $0$. So we expand (\ref{E50}) near $w=0$ and keep the  $w$ term. This gives
\[p^2(1-p)^{\Cs}(1+\Cs \ln(1-p)+\ln(1-p))=0.\]
Setting $p=1-\xi,\Cs=-1+\eta$, we obtain
\[(-1+\xi)^2\xi^{-1+\eta}(1+\eta\ln(\xi))=0,\]
hence
\bals
\eta(\xi)& \sim -1/\ln(\xi),\xi\ra 0,\\
\xi(\eta)&\sim \exp(-1/\eta),\eta \ra 0.
\end{align*}
For instance, for $\Cs=-1+0.09$ we have ($\hat{x}$ always denotes  some solution of (\ref{E50}))  \\
$\hat{\xi}=0.00001494533852483\ldots$ and
$ \eta(\hat{\xi})= 0.09000000000002\ldots,\xi(0.09)=0.00001494533852478\ldots$.

\item  on the diagonal $p'=p$, we set $p=p'=1/2-\eps,\Cs=-1+\eta$. From (\ref{E70}), expand w.r.t. $\xi$, we obtain
 
\[\Cs\sim \frac{ \ln(-16\ln(2)-8\ln(\eps))+\ln(\eps)}{-2\ln(2)-\ln(\eps)}\sim -1-\frac{\ln(2)+\ln(-\ln(\eps))}{\ln(\eps)},\]
hence
\[\eta(\xi)\sim -\frac{\ln(2)+\ln(-\ln(\eps))}{\ln(\eps)},\eps\ra 0.\]
To obtain $\eps$ as a function of $\eta$, we set  $A:=-\ln(\eps)$. We derive, to first order,
\bals
&\ln(2)+\ln(A)-\eta A=0,\\
&A\exp(-\eta A)=1/2,\\
&-\eta A \exp(-\eta A)=-\eta/2,\\
&-\eta A=W_{-1}(-\eta/2),\\
&A=-W_{-1}(-\eta/2)/\eta, \mbox{ for } -\eta/2>-1/e=-0.3678794411\ldots,\\
&\eps(\eta) \sim \exp(W_{-1}(-\eta/2)/\eta),\eta\ra 0,
\end{align*}
where $W(x)$ is the Lambert-W function and the lower branch has $W \leq -1$ and is denoted by $ W_{-1}(x)$. It decreases from
 $W_{-1}(-1/e) = -1$ to $ W_{-1}(0^{−}) = -\infty$.  For instance, for $\eps=10^{-20},\hat{\eta}=0.1005777569\ldots$ and
$\eta(\hat{\eps})=.09821378400137\ldots, \eps(\hat{\eta})= 4.10^{-20}$.

Now $W_{-1}(x)\sim \ln(-x),x\uparrow 0$. Hence
\[\eps(\eta) \sim \exp(\ln(\eta/2)/\eta),\eta\ra 0.\]

\item in the neighbourhood of $p'=1-p$, we set $p'=1-p-\de,\Cs=-1+\eta$. Hence $\qt=\de,q'=p-\de$. As $\de \ra 0$,
 we have $p'\sim 1-p,q'\sim p$. So we expand (\ref{E50}) to first order. We obtain

\[C_1\de^\eta+C_2 q^\eta+C_3(p-\de)^{-1+\eta}=0,\]
with
\[C_1=C_4+C_5\ln(\de),C_4=(p^2+(1-p)^2)\ln(p),C_5=-p^2-(1-p)^2 ,
C_2=-p^2(-\ln(q)+\ln(p)),C_3=-(1-p)^2 p.\]
This leads to
\[\eta(\de)\sim \frac{\ln(C_6)-\ln(\ln(\de))}{\ln(\de)},\de\ra 0,\]
\[C_6=\frac{-1+\ln(1/pq)p^3-2\ln(1/pq)p^4+\ln(1/pq)p^5}{p(2p^2+1-2p)(p-1)^2}.\]
Setting $B:=-\ln(\de),C_7=-C_6$, this leads to
\bals
&Be^{-\eta B}\sim C_7,\\
&-\eta Be^{-\eta B}\sim -\eta C_7,\\
&-\eta B\sim W_{-1}(-\eta C_7),\\
&B\sim -W_{-1}(-\eta C_7)/\eta,\\
&\de(\eta)\sim \exp(\ln(\eta C_7)/\eta),\eta\ra 0,\\
&\eta(\de)\sim (\ln(B)-\ln(C_7))/B,\de\ra 0.
\end{align*}
For instance, for $p=0.75,\eta =0.035$, we obtain $\hat{p'}=0.25-0.161018555971\ldots 10^{-60},
\hat{\de}=.161018555971\ldots10^{-60},C_6=-35.12208439\ldots$ 
and $\eta(\hat{\de})=.009877595163\ldots$. $-\ln(\de)=139\ldots$ is not large  enough, compared with $C_7=34\ldots$ in order to use  $\eta(\hat{\de})$. However, $\ln(B)/B=.03530119866$ which is quite satisfactory. On this other side, $\eta C_7=1.22\ldots$, which is too large ($>1/e$) in our case for allowing using $-W_{-1}(-\eta C_7)/\eta$.
\eit

\subsection*{Acknowledgement.} 
We would like to thank  F.T. Bruss for many illuminating discussions.

\bibliographystyle{plain}

\end{document}